\documentclass[trackchanges,twocolumn]{aastex701}

\usepackage{CJK}

\makeatletter
\newcommand{\autofigwidth}{%
  \if@twocolumn
    \linewidth
  \else
    0.5\linewidth
  \fi
}
\makeatother

\begin{document}
\begin{CJK*}{UTF8}{ipxm}

\title{A Diagnostic Method for Proto-Neutron Star Magnetic Fields by Supernova Fallback Neutrinos}

\author[0000-0002-0700-2223]{Akihiro Inoue}
\affiliation{Department of Earth Science and Astronomy, The University of Tokyo, Meguro, Tokyo 153-8902, Japan}
\email{inoue-a@g.ecc.u-tokyo.ac.jp}

\author[0000-0002-7443-2215]{Yudai Suwa}
\affiliation{Department of Earth Science and Astronomy, The University of Tokyo, Tokyo 153-8902, Japan}
\affiliation{Center for Gravitational Physics and Quantum Information, Yukawa Institute for Theoretical Physics, Kyoto University, Kyoto 606-8502, Japan}
\email{}

\author[0000-0002-9234-813X]{Ryuichiro Akaho}
\affiliation{Faculty of Science and Engineering, Waseda University, Tokyo 169-8555, Japan}
\email{}

\author[0000-0003-3882-3945]{Shinsuke Takasao}
\affiliation{Humanities and Sciences/Museum Careers, Musashino Art University, Kodaira, Tokyo 187-8505, Japan}
\email{}

\author[0000-0003-4299-8799]{Kazumi Kashiyama}
\affiliation{Astronomical Institute, Tohoku University, Sendai, Miyagi 980-8578, Japan}
\affiliation{Kavli Institute for the Physics and Mathematics of the Universe, The University of Tokyo, Kashiwa, Chiba 277-8583, Japan}
\email{}

\author[0000-0003-0114-5378]{Hiroyuki R. Takahashi}
\affiliation{Department of Natural Sciences, Faculty of Arts and Sciences, Komazawa University, Setagaya, Tokyo 154-8525, Japan}
\email{}


\begin{abstract}

Young isolated neutron stars exhibit a wide range of magnetic field strengths.
To understand the origin of this diversity, it is important to observationally constrain magnetic fields during the proto-neutron star (PNS) phase.
We propose a novel diagnostic of PNS magnetic fields using neutrinos associated with supernova fallback.
By performing one-dimensional general-relativistic magnetohydrodynamic simulations of the supernova fallback onto a magnetized PNS, we investigate how the energy spectra of the fallback neutrinos depend on the PNS magnetic field strength and the fallback accretion rate.
We find that stronger magnetic fields and lower accretion rates produce softer spectra because the fallback neutrinos are emitted mainly near the magnetospheric radius, which becomes larger under these conditions.
To quantify this spectral softening, we define a hardness ratio based on event rates above and below $30~{\rm MeV}$ and introduce a hardness-event-rate diagram.
Using this diagram, we identify the parameter region in which our diagnostic is applicable.
Its applicability depends on the PNS magnetic field strength, fallback accretion rate, onset time of the fallback phase, and neutrino mass ordering.
We also investigate the dependence on the PNS radius and find that a larger radius produces softer spectra, indicating a degeneracy between the magnetic field and PNS compactness.
An independent constraint on the PNS compactness would therefore enable us to better constrain the magnetic field.
These results suggest that future multimessenger observations of galactic supernovae can provide a new window into the NS diversity.

\end{abstract}

\keywords{Neutron stars (1108) --- General relativity (641) --- Accretion (14) --- High energy astrophysics (739)}

\section{Introduction} \label{sec:intro}

\par
Young isolated neutron stars (NSs) in the Milky Way, with typical ages of $\lesssim 10$ kyr, are categorized as magnetars, radio pulsars, and central compact objects (CCOs) \citep{Kaspi2010,Enoto2019,Borghese2023}.
These populations are believed to be powered by magnetic, rotational, and thermal energy, respectively.
Their characteristic surface dipole magnetic field strengths differ considerably: $>10^{14}~{\rm G}$ in magnetars, $10^{12\text{--}13}~{\rm G}$ in radio pulsars, and $<10^{11}~{\rm G}$ in CCOs.
However, the origin of such diversity remains an open question.

\par
Various mechanisms have been proposed to explain the magnetic fields of newborn NSs \citep[for a review, see][]{Igoshev2021}.
During the formation of a proto-neutron star (PNS) in a core-collapse supernova, the magnetic field of the progenitor core can be amplified through magnetic flux conservation \citep{Woltjer1964}, turbulent dynamo action \citep{Duncan1992,Thompson1993,Raynaud2020,Masada2022}, magnetorotational instability \citep{Akiyama2003,Obergaulinger2009,Masada2015,Reboul2021}, and the standing accretion shock instability in the post-bounce supernova environment \citep{Blondin2003,Endeve2010,Endeve2012,Foglizzo2015}.
The amplified magnetic fields undergo relaxation and large-scale rearrangement during the fluid PNS phase, and the external dipole field freezes out once a solid crust forms as the PNS cools \citep{Suwa2026}.
Constraining the magnetic field during the PNS phase is therefore essential for understanding how the NS diversity is established.

\par
Neutrinos provide a promising probe of the PNS.
Since neutrinos interact weakly with matter, they can escape from the dense stellar material and carry information from regions close to the PNS.
The present study focuses on neutrinos produced by supernova fallback.
\citet{Akaho2024} performed one-dimensional general relativistic neutrino-radiation hydrodynamic simulations of supernova fallback onto a non-magnetized PNS.
They showed that the fallback neutrinos can be detected by Super-Kamiokande (hereafter, SK) when the fallback accretion rate is $\gtrsim10^{-3}~M_\odot~{\rm s}^{-1}$ for a source distance of $10~{\rm kpc}$.
The fallback neutrinos with energies of $\gtrsim 30~{\rm MeV}$ are expected to appear as a high-energy tail in the energy spectrum.
These results motivate us to investigate the fallback neutrinos as a probe of the PNS magnetic field.

\par
The PNS magnetic field affects the dynamics of the supernova fallback and the neutrino spectra.
In our previous work \citep{Inoue2026}, we conducted one-dimensional general relativistic magnetohydrodynamic (GRMHD) simulations to investigate the dynamics of the supernova fallback onto a magnetized PNS \citep[see also][]{Bernal2010,Bernal2013}.
We demonstrated that the system is divided into three regions \citep{Torres2016}: the freefall region, the post-shock region, and the PNS magnetosphere.
The boundaries between these regions are characterized by the shock and magnetospheric radii, respectively.
The fallback neutrinos are emitted primarily at the magnetospheric radius.
This radius is larger for stronger magnetic fields and lower accretion rates.
The larger magnetospheric radius leads to a lower gas temperature there.
Therefore, the neutrino spectra of the fallback accretion should encode information about the PNS magnetic field.

\par
To test this idea, we perform one-dimensional GRMHD simulations and investigate how the spectral properties of the fallback neutrinos depend on the PNS magnetic field strength, the fallback accretion rate, and the PNS radius.
Our simulations incorporate neutrino cooling and a realistic equation of state (EoS).
The latter was not included in \citet{Inoue2026}.
We also perform two-dimensional simulations to assess the impact of multidimensional effects.

\par
This paper is organized as follows.
We describe our numerical method in Section~\ref{sec:method} and present the numerical results in Section~\ref{sec:results}.
Observational implications and multidimensional effects are discussed in Sections~\ref{sec:implications} and \ref{sec:multiD}, respectively.
Finally, we summarize our findings in Section~\ref{sec:conclusion}.
We also discuss model limitations in Section \ref{sec:limitations}.
Hereafter, the speed of light $c$ and the gravitational constant $G$ are normalized to unity unless otherwise specified.
We denote the mass and radius of the PNS by $M_{\rm PNS}$ and $r_{\rm PNS}$, respectively.

{\section{Numerical Method} \label{sec:method}}

\par
We numerically solve the GRMHD equations with neutrino cooling in Schwarzschild polar coordinates $(t,r,\theta,\phi)$.
The simulations are performed in one dimension under the assumption of spherical symmetry.
We use the numerical code {\tt UWABAMI} \citep[e.g.,][]{Takahashi2017}.
Throughout this paper, Greek indices denote space–time components, while Latin indices represent spatial components.

{\subsection{Basic equations}\label{sec:basic_equations}}

\par
The basic equations of GRMHD are as follows:
\begin{eqnarray}
    \nabla_\mu\left(\rho u^\mu\right)&=&0,
    \label{eq:mass_cons}
    \\
    \nabla_\mu\left(T^{\mu\nu}\right)&=&\sqrt{-g}Q^\nu,
    \label{eq:momentum_cons_gas}
    \\
    \partial_t\left(\sqrt{-g}B^i\right)&=&-\partial_j\left\{\sqrt{-g}\left(b^i u^j-b^j u^i\right)\right\},
    \label{eq:induction_eq}
\end{eqnarray}
where $\rho$ is the gas density, $u^\mu$ is the four-velocity of the gas, $B^i$ is the magnetic field vector in the laboratory frame, $b^\mu$ is the magnetic four-vector in the fluid frame, and $g={\rm det}(g_{\mu\nu})$ is the determinant of the metric.
The energy-momentum tensor of ideal MHD is expressed as
\begin{eqnarray}
    {T}^{\mu\nu}&=&
    \left(\rho+e_{\rm tot}+p_{\rm tot}+e_{\rm mag}+p_{\rm mag}\right)u^{\mu}u^{\nu}\nonumber\\
    &+&\left(p_{\rm tot}+p_{\rm mag}\right)g^{\mu\nu}
    -b^\mu b^\nu,
    \label{eq:Tmunu_gas}
\end{eqnarray}
where $e_{\rm tot}$ is the internal energy density of the fluid, $p_{\rm tot}$ is the fluid pressure, $e_{\rm mag}=b^\mu b_\mu/2$ is the magnetic energy density, and $p_{\rm mag}=b^\mu b_\mu/2$ is the magnetic pressure.
We write the source term accounting for neutrino emission as $Q^\mu=-\dot{q}u^\mu$, where $\dot{q}$ is the cooling rate per unit volume and consists of contributions from pair neutrino emission $\dot{q}_{\rm pair}$ ($e^-+e^+\rightarrow \nu+\bar{\nu}$) and lepton capture processes $\dot{q}_{\rm LCP}$ ($p+e^-\rightarrow n+\nu_{\rm e}$ and $n+e^+\rightarrow p+\bar{\nu}_{\rm e}$) \citep{Itoh1989,Qian1996}.
The pair process includes contributions from all neutrino flavors by setting the number of neutrino flavors other than the electron flavor to $n=2$ \citep[see][for details]{Itoh1989}.
The contributions of the other processes discussed in \citet{Itoh1989,Itoh1996} are subdominant compared to $\dot{q}_{\rm pair}$ and $\dot{q}_{\rm LCP}$, and therefore we ignore them.

\par
We adopt the Helmholtz equation of state (EoS) \citep{Timmes2000}, which is based on a table interpolation of the Helmholtz free energy described by \citet{Timmes1999}.
The Helmholtz EoS includes contributions from blackbody radiation ($e_{\rm rad}$ and $p_{\rm rad}$), fully ionized nuclei ($e_{\rm ion}$ and $p_{\rm ion}$), degenerate and non-degenerate relativistic electrons ($e_{\rm ele}$ and $p_{\rm ele}$) and positrons ($e_{\rm pos}$ and $p_{\rm pos}$), and Coulomb corrections ($e_{\rm Cou}$ and $p_{\rm Cou}$).
The internal energy density and pressure of the fluid are obtained by summing the contributions of these components:
\begin{eqnarray}
    e_{\rm tot}&=&e_{\rm rad}+e_{\rm ion}+e_{\rm ele}+e_{\rm pos}+e_{\rm Cou},\\
    p_{\rm tot}&=&p_{\rm rad}+p_{\rm ion}+p_{\rm ele}+p_{\rm pos}+p_{\rm Cou}.
    \label{eq:ptot_EOS}
\end{eqnarray}
The fluid is assumed to consist of free protons, neutrons, and electrons, with a constant electron fraction $Y_{\rm e}=0.5$ for simplicity.
Given a temperature $T$, a density $\rho$, and an electron fraction $Y_{\rm e}$, the Helmholtz EoS package returns the corresponding $e_{\rm tot}$ and $p_{\rm tot}$.
Therefore, we adopt the gas temperature as the primitive variable instead of the pressure.

{\subsection{Model description}\label{sec:numrical_model}}

\par
We describe the supernova fallback as a system in which non-magnetized material accretes onto a magnetized PNS in a direction perpendicular to its dipole magnetic field \citep{Cumming2001,Inoue2026}.
The numerical setup and schemes are explained below.

\subsubsection{Initial condition}

\par
The PNS magnetic field is parameterized by the magnetic field strength at $r=10~{\rm km}$, $B_{\rm 10km}$.
We keep $B_{\rm 10km}$ fixed across models with different $r_{\rm PNS}$ to isolate the effect of $r_{\rm PNS}$.
The initial magnetic field is given by $B^{(\theta)}=B_{\rm 10km}(r/10~{\rm km})^{-3}$ and $B^r=B^\phi=0$.
The parentheses in the indices of physical quantities represent the values in the orthonormal frame of a static observer.
We also investigate the non-magnetized case ($B_{\rm 10km}=0$) for comparison.

\par
We initially set the atmosphere in MHD equilibrium to maintain the radial profile of $B^{(\theta)}$.
From the relation of $dp_{\rm mag}/dr=-M_{\rm PNS}\rho/r^2$, the initial gas density is given by $\rho=(3/4\pi)B_{\rm 10km}^2(r_{\rm PNS}/r_{\rm g})(r/r_{\rm PNS})^{-5}$ with $r_{\rm g}=M_{\rm PNS}$.
We set the initial pressure ratio to $p_{\rm tot}/p_{\rm mag}=0.1$.
For the non-magnetized case, we set the hydrostatic atmosphere that satisfies $dp_{\rm tot}/dr=-M_{\rm PNS}\rho/r^2$.
The atmosphere is assumed to be described by a polytropic relation of $p_{\rm tot}\propto \rho^\Gamma$ with $\Gamma=4/3$.
Defining $\rho_{\rm out}$ as the gas density at the outer boundary, this relation gives us $\rho=\rho_{\rm out}(r/r_{\rm PNS})^{-1/(\Gamma-1)}$ and $p_{\rm tot}=\rho_{\rm out}[(\Gamma-1)/\Gamma](r_{\rm g}/r_{\rm PNS})(r/r_{\rm PNS})^{-\Gamma/(\Gamma-1)}$.
The initial gas temperature is calculated from the given $p_{\rm tot}$ and $\rho$ using the Newton-Raphson method.
In all models, we initially set $u^i=0$.
This initial atmosphere does not affect the neutrino luminosity and spectra.

\subsubsection{Boundary condition\label{sec:BC}}

\par
The fallback phase begins several seconds or more after core bounce \citep[see e.g.,][]{Fryer2009,Janka2022,Wang2023,Wang2024,Shinoda2025}.
The fallback accretion rate depends on time $t$ and can be approximately expressed as \citep{Michel1988,Chevalier1989,Metzger2018,Barrere2022}
\begin{eqnarray}
    \dot{M}_{\rm fb}(t)
    =
    \frac{2}{3} \frac{M_{\rm fb}}{t_{\rm fb}}
    \left(\frac{1}{1+t/t_{\rm fb}}\right)^{5/3},
    \label{eq:fallback}
\end{eqnarray}
where we write the fallback mass and time scale as $M_{\rm fb}$ and $t_{\rm fb}$, respectively.
These quantities depend on the progenitor structure \citep[see e.g.,][]{Ugliano2012,Ertl2016a,Ertl2016b,Sawada2022}.
Their typical values are roughly $M_{\rm fb}\sim 10^{-4}-10^{-1}~{M_\odot}$ and $t_{\rm fb}\sim 10^0-10^3~{\rm s}$, respectively.
Therefore, $\dot{M}_{\rm fb}$ ranges from $10^{-7}$ to $10^{-1}~{ M_\odot~\rm s^{-1}}$.
For simplicity, we assume a constant $\dot{M}_{\rm fb}$.
This assumption corresponds to the limit $t\ll t_{\rm fb}$ in Equation (\ref{eq:fallback}).

\par
From the outer boundary, we inject the non-magnetized material with a freefall velocity $v_{\rm ff}$.
We compute the temperature using the Newton–Raphson method such that the Mach number is equal to 10.
Although the Mach number of the fallback accretion flow is poorly constrained, we have confirmed that our results are insensitive to its exact value as long as the inflow is sufficiently supersonic.

\par
For the inner boundary, we impose the following conditions at the PNS surface.
The gas velocity is set to zero in the ghost cells.
We fix $B^i$ to its initial value and impose a zero-gradient boundary condition on $\rho$ and $T$ in these cells.
Since the electrical conductivity inside the PNS is expected to be high \citep{Igoshev2021}, the electric field can be neglected. 
We therefore set the numerical flux of the radial component in Equation (\ref{eq:induction_eq}) to zero at $r = r_{\rm PNS}$.

\subsection{Model parameters\label{sec:model_param}}

\par
Table \ref{tab:table1} presents the model parameters adopted in the present study.
We investigate five values of $B_{\rm 10km}$: $0$, $3\times10^{14}$, $10^{15}$, $3\times10^{15}$, and $10^{16}~{\rm G}$.
Following \citet{Akaho2024}, we focus on relatively high fallback accretion rates in the range
$10^{-3}~M_\odot~{\rm s^{-1}}\le \dot{M}_{\rm fb}\le10^{-2}~M_\odot~{\rm s^{-1}}$.
The rationale for the adopted values of the outer-boundary radius, $r_{\rm out}$, is described in Section \ref{sec:numerical_scheme}.

\par
Typical values of the PNS mass and radius are adopted in our simulations.
The PNS cools through neutrino emission after its birth and contracts.
This contraction phase is considered to last within $\sim 10~{\rm s}$ \citep[][]{Fischer2010,Suwa2014}.
The mass and radius of the PNS are typically $\approx 1-2~M_\odot$ and $\approx 10$–$20~{\rm km}$, respectively \citep[e.g.,][]{Fischer2010,Nagakura2022}.
They are related through the nuclear EoS and the thermodynamic state of the PNS, but we fix $M_{\rm PNS}=1.4~M_\odot$ in all models and treat $r_{\rm PNS}$ as an independent model parameter.
We investigate four cases with $r_{\rm PNS}=10$, $13$, $16$, and $20~{\rm km}$, and keep it fixed in the simulations for simplicity.
We first present the results for $r_{\rm PNS}=10~{\rm km}$, while the results for the other cases are provided in Section \ref{sec:PNS_radius}.

\begin{deluxetable}{cccccc}[htb]
\tablecaption{Parameters for numerical models\label{tab:table1}}
\tablehead{
\colhead{Parameters}
&\colhead{$B_{\rm 10km}$}
&\colhead{$\dot{M}_{\rm fb}$}
&\colhead{$r_{\rm out}$}
&\colhead{$r_{\rm PNS}$}\\
\colhead{Unit}
&\colhead{$[\rm G]$}
&\colhead{$[M_\odot~\rm s^{-1}]$}
&\colhead{$[{\rm km}]$}
&\colhead{$[{\rm km}]$}
}
\startdata
{\tt BnM1em3R10}    & $0$ & $10^{-3}$ & $10^2$ & 10\\
{\tt BnM2em3R10}    & $0$ & $2\times10^{-3}$ & $10^2$ &10\\
{\tt BnM5em3R10}    & $0$ & $5\times10^{-3}$ & $10^2$ & 10\\
{\tt BnM1em2R10}    & $0$ & $10^{-2}$ & $10^2$ & 10\\
{\tt BnM1em3R13}    & $0$ & $10^{-3}$ & $10^3$ & 13\\
{\tt BnM1em3R16}    & $0$ & $10^{-3}$ & $10^3$ & 16\\
{\tt BnM1em3R20}    & $0$ & $10^{-3}$ & $10^3$ & 20\\
{\tt B3e14M1em3R10} & $3\times10^{14}$ & $10^{-3}$ & $10^3$ & 10\\
{\tt B3e14M2em3R10} & $3\times10^{14}$ & $2\times10^{-3}$ & $10^2$ & 10\\
{\tt B3e14M5em3R10} & $3\times10^{14}$ & $5\times10^{-3}$ & $10^2$ & 10\\
{\tt B3e14M1em2R10} & $3\times10^{14}$ & $10^{-2}$ & $10^2$ & 10\\
{\tt B3e14M1em3R13} & $3\times10^{14}$ & $10^{-3}$ & $10^3$ & 13\\
{\tt B3e14M1em3R16} & $3\times10^{14}$ & $10^{-3}$ & $10^3$ & 16\\
{\tt B3e14M1em3R20} & $3\times10^{14}$ & $10^{-3}$ & $10^3$ & 20\\
{\tt B1e15M1em3R10} & $10^{15}$ & $10^{-3}$ & $10^3$ & 10\\
{\tt B1e15M2em3R10} & $10^{15}$ & $2\times10^{-3}$ & $10^3$ & 10\\
{\tt B1e15M5em3R10} & $10^{15}$ & $5\times10^{-3}$ & $10^2$ & 10\\
{\tt B1e15M1em2R10} & $10^{15}$ & $10^{-2}$ & $10^2$ & 10\\
{\tt B1e15M1em3R13} & $10^{15}$ & $10^{-3}$ & $10^3$ & 13\\
{\tt B1e15M1em3R16} & $10^{15}$ & $10^{-3}$ & $10^3$ & 16\\
{\tt B1e15M1em3R20} & $10^{15}$ & $10^{-3}$ & $10^3$ & 20\\
{\tt B3e15M1em3R10} & $3\times10^{15}$ & $10^{-3}$ & $10^3$ & 10\\
{\tt B3e15M2em3R10} & $3\times10^{15}$ & $2\times10^{-3}$ & $10^3$ & 10\\
{\tt B3e15M5em3R10} & $3\times10^{15}$ & $5\times10^{-3}$ & $10^3$ & 10\\
{\tt B3e15M1em2R10} & $3\times10^{15}$ & $10^{-2}$ & $10^3$ & 10\\
{\tt B3e15M1em3R13} & $3\times10^{15}$ & $10^{-3}$ & $10^4$ & 13 \\
{\tt B3e15M1em3R16} & $3\times10^{15}$ & $10^{-3}$ & $10^4$ & 16\\
{\tt B3e15M1em3R20} & $3\times10^{15}$ & $10^{-3}$ & $10^4$ & 20\\
{\tt B1e16M2em3R10} & $10^{16}$ & $2\times10^{-3}$ & $10^4$ & 10\\
{\tt B1e16M5em3R10} & $10^{16}$ & $5\times10^{-3}$ & $10^4$ & 10\\
{\tt B1e16M1em2R10} & $10^{16}$ & $10^{-2}$ & $10^3$ & 10\\
\enddata
\tablecomments{
The model names are shown in the first column: ``{\tt BXeYY}'' means the magnetic field strength at $r=10~{\rm km}$ of ${\rm X}\times10^{{\rm YY}}~{\rm G}$ (``{\tt Bn}'' is the case of $B_{\rm 10km}=0$), ``{\tt MXemY}'' denotes the fallback mass accretion rate of ${\rm X}\times10^{-{\rm Y}}~{\rm M_\odot~s^{-1}}$, and ``{\tt RXX}'' represents the PNS radius of ${\rm XX}~{\rm km}$.
We present the magnetic field strength at $r=10~{\rm km}$ ($B_{\rm 10km}$), the mass accretion rate at $r=r_{\rm out}$ ($\dot{M}_{\rm fb}=-4\pi r_{\rm out}^2\rho_{\rm out}u^r$), the radius of the outer boundary ($r_{\rm out}$), and the PNS radius ($r_{\rm PNS}$).
We adopt model {\tt B1e15M1em3R10} as our fiducial model.}
\end{deluxetable}

\subsection{Numerical schemes\label{sec:numerical_scheme}}

\par
The numerical methods used to solve the basic equations in Section \ref{sec:basic_equations} are summarized below.
For spatial reconstruction, we use the harmonic mean proposed by \citet{vanLeer1977}, and evaluate the numerical fluxes with the Harten–Lax–van Leer \citep[HLL;][]{Harten1983} solver.
\footnote{
The HLL solver is more numerically diffusive than HLLC/HLLD \citep{Miyoshi2005,Mignone2006b,Mignone2009,Mattia2022}.
In the present simulations, the effect of such numerical dissipation is particularly relevant near the magnetospheric boundary, where the neutrino emission is significant.
To check how the numerical dissipation affects our conclusion, we perform a resolution study with $N_r=8192$ and $11586$ for our fiducial model and compare them with the original calculation at $N_r=16384$.
The spectra show small differences between $N_r=8192$, $11586$ and $16384$ and the quantitative convergence is not established. 
Therefore, comparisons with less diffusive Riemann solvers and further resolution tests are needed to establish quantitative convergence.
}
Time integration of the advection terms is performed using an explicit second-order Runge–Kutta method.
To recover the primitive variables from the conserved ones, we employ the inversion scheme proposed by \citet{Noble2006}.
This scheme involves an additional inversion step from specific enthalpy to temperature \citep[][]{Siegel2018}, which is performed using the Newton-Raphson method.
If this inversion scheme fails to recover the primitive variables, we attempt to recover them using an entropy-conserving approach.
If both attempts fail, we use the arithmetic mean of the primitive variables from neighboring cells where the inversion has successfully converged.

\par
The source term associated with neutrino cooling, $Q^\mu$, is implicitly integrated using the Newton–Raphson method.
This implicit integration is computationally expensive because it requires simultaneously solving for four variables, $T$ and $u^i$.
We therefore adopt the following approximation.
The source term of $Q^\mu$ in the orthonormal frame satisfies $Q^{(i)}/Q^{(t)}=\mathcal{O}(v/c)$, where $v$ is the gas speed.
Since $v$ is much smaller than $c$ in the region where neutrino cooling is significant, we neglect spatial components by setting $Q^i=0$.
This approximation enables us to solve the source term using the Newton-Raphson method for a single variable $T$, thereby reducing computational cost.
Although we do not include energy and momentum transport due to neutrino-matter interactions, these processes do not affect our conclusions within the density range treated in the present study.

\par
To maintain numerical stability, we impose floor values on the density and pressure: $\rho_{\rm fl}=10^{-4}\rho_{\rm out}(r/r_{\rm PNS})^{-3}$ and $p_{\rm fl}=10^{-6}\rho_{\rm out}(\Gamma - 1)(r/r_{\rm PNS})^{-4}$.
In regions where the kinetic or magnetic energy is much greater than the thermal energy, numerical errors can result in a negative pressure.
These floors help us avoid the problem.

\par
The computational domain and its resolution are described below.
The domain extends from $r_{\rm PNS}$ to $r_{\rm out}$.
In models with a low $\dot{M}_{\rm fb}$, a strong $B_{\rm 10km}$, or a large $r_{\rm PNS}$, the post-shock region extends to a large radius \citep{Chevalier1989,Houck1991,Bernal2010,Akaho2024,Inoue2026}.
We therefore adopt a large $r_{\rm out}$ for these models.
As described in Section \ref{sec:results}, a thin magnetosphere forms in our simulations.
To resolve the magnetosphere with sufficient accuracy, we adopt a radial resolution of $N_r=16384$.
The radial grid is logarithmically spaced, such that the grid radius increases exponentially as $e^{x_a}$ with $x_a=\ln r_{\rm PNS}+a\times dX$, where $a$ is the radial grid index, and $dX=(\ln r_{\rm out}-\ln r_{\rm PNS})/N_r$.

{\section{Results}\label{sec:results}}

\par
We first describe the accretion structure and time evolution of the system in Sections \ref{sec:accr_struc} and \ref{sec:time_evol}, respectively.
In these sections, we focus on our fiducial model {\tt B1e15M1em3R10} because the qualitative features are similar among the models with $r=10~{\rm km}$.
We examine the energy spectra of the fallback neutrinos in Section \ref{sec:spectra} and introduce our diagnostic method for the PNS magnetic field in Section \ref{sec:HED}.
Throughout these sections, we present the results for $r_{\rm PNS}=10~{\rm km}$, while the dependence on $r_{\rm PNS}$ is discussed in Section \ref{sec:PNS_radius}.

{\subsection{Accretion structure}\label{sec:accr_struc}}

\begin{figure}[tb]
\centering
\includegraphics[width=\autofigwidth]{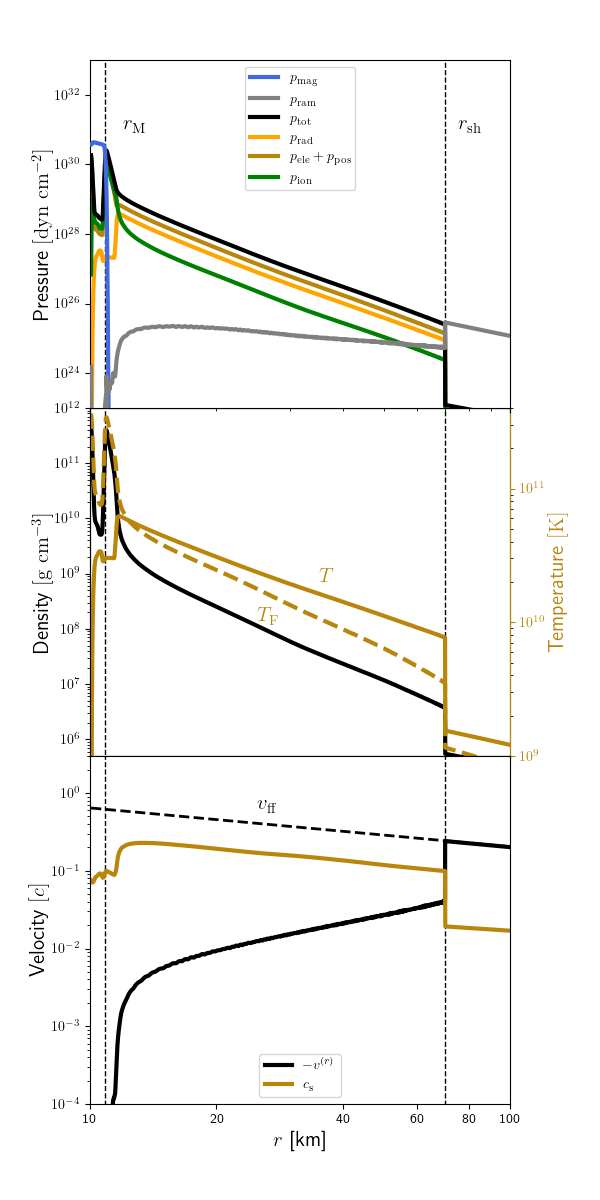}
\caption{
Radial profiles of the physical quantities in our fiducial model {\tt B1e15M1em3R10} at $t\approx 0.17~{\rm s}$.
The vertical dashed lines indicate the magnetospheric and shock radii, denoted by $r_{\rm M}$ and $r_{\rm sh}$, respectively.
Top panel: radial profiles of pressures.
Middle panel: gas density and temperature.
The dashed line represents the Fermi temperature $T_{\rm F}$.
Bottom panel: infalling velocity and the sound speed.
The dashed line represents the freefall velocity.
\label{fig:figure1}}
\end{figure}

\par
The top panel of Figure \ref{fig:figure1} shows radial profiles of various pressures at $t \approx 0.17~{\rm s}$.
We write the ram pressure as $p_{\rm ram}=\rho (v^{(r)})^2$ with $v^{(r)}=u^{(r)}/u^{(t)}$.
The solid lines represent $p_{\rm mag}$ (blue), $p_{\rm tot}$ (black), and $p_{\rm ram}$ (gray).
The total fluid pressure consists of the contributions described in Equation (\ref{eq:ptot_EOS}), which are also presented.
Since $p_{\rm Cou}$ is much smaller than the other contributions, we do not show it in this panel.
The vertical dashed lines indicate two characteristic radii: the magnetospheric radius $r_{\rm M}$ and the shock radius $r_{\rm sh}$.
We define $r_{\rm M}$ and $r_{\rm sh}$ as the radii where $p_{\rm mag}=p_{\rm tot}$ and $p_{\rm tot}=p_{\rm ram}$, respectively.
These radii divide the system into three regions: the freefall region for $r>r_{\rm sh}$, the post-shock region for $r_{\rm M}<r<r_{\rm sh}$, and the PNS magnetosphere for $r_{\rm PNS}<r<r_{\rm M}$.

\par
The magnetospheric radius lies very close to the PNS surface, $r_{\rm M}\approx r_{\rm PNS}$, indicating that the fallback material compresses the PNS magnetic field and forms a thin magnetosphere.
The compressed magnetic field decelerates the infalling material and drives an accretion shock.
We write the shock formation time as $t_{\rm shock}$, which is defined as the moment when a region satisfying
$p_{\rm tot}>\max(p_{\rm mag},p_{\rm ram})$ and $p_{\rm rad}>10^{24}~{\rm dyn~cm^{-2}}$ first appears.
The shock formation time is approximately consistent with the freefall time from the outer boundary to the PNS surface.

\par
The fluid pressure in the post-shock region is dominated by the contributions of relativistic, non-degenerate electrons and positrons \citep{Houck1991}.
The combined pressure of electrons and positrons is well described by the relativistic, non-degenerate limit $p_{\rm ele}+p_{\rm pos}\approx (7/4)p_{\rm rad}$.
However, the combined pressure deviates from this limit near $r\approx r_{\rm M}$, and $p_{\rm ion}$ becomes comparable to $p_{\rm ele}+p_{\rm pos}$.
These features near $r=r_{\rm M}$ can be understood from the thermodynamic structure, as described below.

\par
The middle panel of Figure~\ref{fig:figure1} shows the radial profiles of $\rho$ and $T$.
The dashed line represents the Fermi temperature.
When the gas temperature is lower than this temperature, the degenerate pressure of electrons becomes important.
With the Fermi momentum $p_{\rm F}$, electron mass $m_{\rm e}$, and Boltzmann constant $k$, the Fermi temperature is expressed as
\begin{eqnarray}
    T_{\rm F}
    =\frac{\sqrt{p_{\rm F}^2+m_{\rm e}^2}-m_{\rm e}}{k}.
\end{eqnarray}
Both $\rho$ and $T$ generally increase inward in the post-shock region.
However, $T$ drops sharply near $r=r_{\rm M}$.
The sharp decrease in $T$ is caused by neutrino cooling.
Since neutrino cooling is significant at $r\approx r_{\rm M}$, the fluid pressure decreases there.
The decrease in the fluid pressure results in an increase in $\rho$.
The decrease in $T$ and the increase in $\rho$ produce $T<T_{\rm F}$ near $r\approx r_{\rm M}$.
The electrons are therefore partially degenerate, and the positron abundance is strongly suppressed relative to the electron abundance.
Consequently, the combined electron and positron pressure is dominated by the contribution of relativistic, degenerate electrons, $p_{\rm ele}=p^4_{\rm F}/(12\pi^2\hbar^3)$.
In addition, the increase in $\rho$ also leads to an increase in $p_{\rm ion}$.
We obtain $p_{\rm ele}\sim p_{\rm ion}$ for the fiducial values of $\rho=10^{11}~{\rm g cm^{-3}}$ and $T=10^{10}~{\rm K}$.

\par
The thermodynamic features explained above indicate that the post-shock flow cannot be accurately described by a simple ideal-gas approximation.
The pressure is determined not only by ions and radiation, but also by relativistic electron-positron pairs and electron degeneracy.
Therefore, a realistic EoS including these contributions is required to accurately calculate the thermodynamic state of the supernova fallback.
The Helmholtz EoS adopted in the present study satisfies this requirement \citep{Bernal2010,Bernal2013,Sakurai2026}.
\footnote{
Although the densities in our simulations remain well below nuclear saturation density, the assumption of free protons and neutrons may affect the thermodynamic properties of the gas.
To assess this effect, we evaluate the total pressure using the nuclear EoS of \citet{Lattimer1991} with a nuclear incompressibility of 220 MeV (LS220) and compare it with that obtained using the Helmholtz EoS.
For this comparison, we adopt the density and temperature profiles in model B1e16M1em2R10.
We find that both EoSs yield similar pressures for $r\gtrsim 14~{\rm km}$, whereas the pressure obtained with LS220 is up to approximately 30\% lower than that obtained with the Helmholtz EoS for $r\lesssim 14~{\rm km}$. 
The reduction in the total pressure is primarily caused by the formation of heavy nuclei in LS220, which reduces the number density of freely moving baryonic particles and consequently decreases the baryonic thermal pressure.
These results demonstrate that nuclear composition can affect the local pressure even at densities well below nuclear saturation density.
Implementing a nuclear EoS in our GRMHD simulation code and assessing its impact on the neutrino spectra (see Sections \ref{sec:spectra} and \ref{sec:HED}) are important future work.
}

\par
The bottom panel of Figure \ref{fig:figure1} shows the radial profiles of the infalling velocity $-v^{(r)}$ and the sound speed $c_{\rm s}$.
The sound speed is calculated from the Helmholtz package.
The dashed line shows the freefall speed.
The flow is supersonic outside the shock radius ($r>r_{\rm sh}$) and subsonic inside it ($r<r_{\rm sh}$), consistent with the presence of an accretion shock.
In the neutrino-emitting region $r\approx r_{\rm M}$, we find $-v^{(r)}\ll c$, supporting our approximation $Q^{(i)}=0$.

{\subsection{Time evolution of the system}\label{sec:time_evol}}

\par
To understand the temporal properties of the system, we examine the shock and magnetospheric radii as functions of time.
We also show the time evolution of the luminosity and mean energy of the fallback neutrinos and compare them with those predicted from a representative PNS cooling model.

\begin{figure}[tb]
\centering
\includegraphics[width=\autofigwidth]{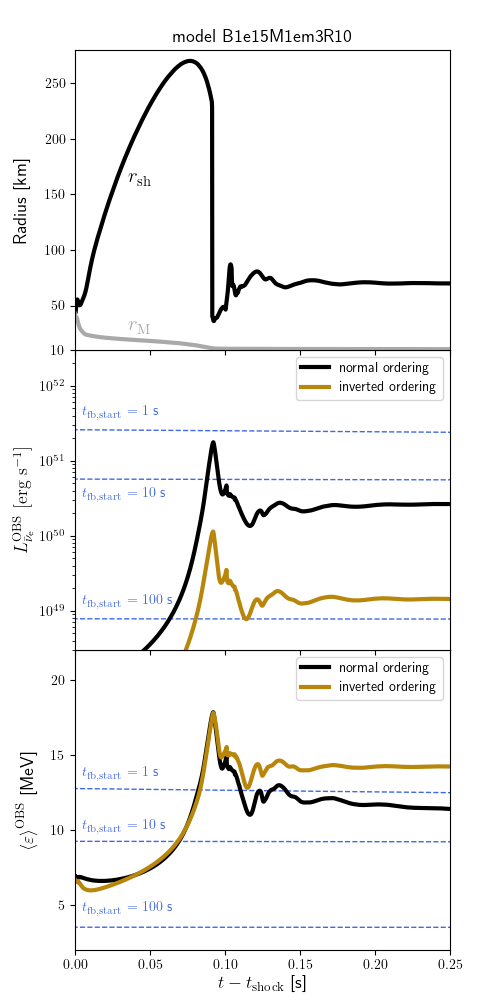}
\caption{
Top panel: time evolution of the shock radius $r_{\rm sh}$ (black) and the magnetospheric radius $r_{\rm M}$ (grey).
Middle panel: neutrino light curves for the normal and inverted orderings.
Bottom panel: the mean neutrino energy for the normal and inverted orderings.
The dashed lines in the middle and bottom panels represent analytic solutions of the luminosity and the mean energy of neutrinos associated with the PNS cooling \citep{Suwa2021} (see text for details).
In all panels, the time is measured relative to the onset of the shock formation.
We adopt the fiducial model {\tt B1e15M1em3R10}.
\label{fig:figure2}}
\end{figure}

\par
The top panel of Figure~\ref{fig:figure2} shows the shock and magnetospheric radii as a function of time measured relative to the onset of the shock formation.
The shock expands for $t-t_{\rm shock}\lesssim 0.07~{\rm s}$ and stalls at $t-t_{\rm shock}\approx 0.08~{\rm s}$.
The radius decreases around $t-t_{\rm shock}\approx 0.1~{\rm s}$ because neutrino cooling becomes effective \citep[see][for details]{Inoue2026}.
Since the neutrino cooling balances the accretion heating, the system relaxes towards a steady state for $t>0.15~{\rm s}$.

\par
The method for calculating the fallback luminosity is described as follows.
The neutrino luminosity consists of contributions from the pair and lepton capture processes (see Section \ref{sec:basic_equations}).
We denote the cooling rate of $e^-+e^+\rightarrow \nu_{\rm e}+\bar{\nu}_{\rm e}$ by $\dot{q}_{\rm pair,e}$ and calculate it by setting $n=0$ \citep{Itoh1989,Itoh1996}.
The cooling rate of $e^-+e^+\rightarrow \nu_{\rm X}+\bar{\nu}_{\rm X}$ is then given by $\dot{q}_{\rm pair,X}=\dot{q}_{\rm pair}-\dot{q}_{\rm pair,e}$.
We introduce the factors $f_{\nu,\rm pair}$ and $f_{\bar{\nu},\rm pair}$ to account for the contributions of $\nu$ and $\bar{\nu}$ to $\dot{q}_{\rm pair}$, respectively.
We take $f_{\nu,\rm pair}=f_{\bar{\nu},\rm pair}=0.5$.
The cooling rate $\dot{q}_{\rm LCP}$ includes contributions from two reactions: $p+e^-\rightarrow n+\nu_{\rm e}$ and $n+e^+\rightarrow p+\bar{\nu}_{\rm e}$.
We assume that the contributions of both reactions are equal.
Therefore, factors accounting for these contributions to $\dot{q}_{\rm LCP}$ are taken as $f_{\nu,\rm LCP}=f_{\bar{\nu},\rm LCP}=0.5$.
From these assumptions, the effective cooling rates for $\bar{\nu}_{\rm e}$ and $\bar{\nu}_{\rm X}$ are given by
\begin{eqnarray}
    \dot{q}^{\rm eff}_{\rm e}&=&f_{\bar{\nu},\rm pair}\dot{q}_{\rm pair,e}
    +f_{\bar{\nu},\rm LCP}\dot{q}_{\rm LCP},\\
    \dot{q}^{\rm eff}_{\rm X}&=&f_{\bar{\nu},\rm pair}\dot{q}_{\rm pair,X},
\end{eqnarray}
respectively.
We compute the luminosities for electron-antineutrinos and heavy-leptonic antineutrinos by integrating these effective cooling rates over the volume:
\begin{eqnarray}
    L_{\bar{\nu}_{\rm e}}^{\rm SRC}
    &=&4\pi\int\dot{q}^{\rm eff}_{\rm e}r^2dr,\label{eq:nu_e}\\
    L_{\bar{\nu}_{\rm X}}^{\rm SRC}
    &=&4\pi\int \dot{q}^{\rm eff}_{\rm X}r^2dr,\label{eq:nu_X}
\end{eqnarray}
respectively.
Throughout this study, we neglect the effect of the gravitational redshift.
\footnote{
For $M_{\rm PNS}=1.4~M_\odot$ and an emission radius of $r_{\rm M}\approx 10-20~{\rm km}$, the gravitational redshift can reduce the neutrino energies measured by a distant observer to a fraction $\alpha=(1-2M_{\rm PNS}/r_{\rm M})^{1/2}\approx 0.77-0.89$ of their local values.
This effect softens the observed neutrino spectra.
A quantitative assessment of its impact on our diagnostic is left for future work.
}

\par
In addition, we consider the effects of neutrino flavor conversions \citep{Dighe2000}.
We follow the same procedure as \citet{Suwa2025}.
In the presence of Mikheyev–Smirnov–Wolfenstein (MSW) flavor conversion, the observed electron-antineutrino luminosity can be written as a linear combination of the electron and heavy-leptonic antineutrino luminosities:
\begin{equation}
L^{\rm OBS}_{\bar{\nu}_{\rm e}}=
\left\{
\begin{array}{ll}
\cos^2\theta_{12}\,L^{\rm SRC}_{\bar{\nu}_{\rm e}}
+\sin^2\theta_{12}\,L^{\rm SRC}_{\bar{\nu}_{\rm X}}
& (\mbox{normal ordering}),\\
L^{\rm SRC}_{\bar{\nu}_{\rm X}}
& (\mbox{inverted ordering}),
\end{array}
\right.
\label{eq:MSW}
\end{equation}
where $\theta_{12}$ is the solar mixing angle, $\cos^2\theta_{12}\approx0.696$ and $\sin^2\theta_{12}\approx 0.304$ \citep{Esteban2020}.

\par
The middle panel of Figure~\ref{fig:figure2} shows $L^{\rm OBS}_{\bar{\nu}_{\rm e}}$ as a function of time for different neutrino mass orderings.
Since the luminosity for the inverted ordering does not include the contributions from the lepton capture processes (see Equations (\ref{eq:nu_e})-(\ref{eq:MSW})), it is lower than that for the normal ordering by about an order of magnitude.
This difference indicates that, in the normal ordering, the contribution from the lepton capture processes dominates over that from the pair process.
The luminosities increase abruptly for $0.05~{\rm s}\lesssim t-t_{\rm shock}\lesssim 0.1~{\rm s}$ and reach their maximum values at $t-t_{\rm shock}\approx 0.1~{\rm s}$.
The luminosities become nearly constant after the system reaches the quasi-steady state ($t>0.15~{\rm s}$).

\par
We find that a time delay arises between the shock formation and the increase in $L^{\rm OBS}_{\bar{\nu}_{\rm e}}$.
The observed neutrinos originate mainly from a region around $r\sim r_{\rm M}$.
The gas density and temperature in this region are not sufficiently high to produce a high luminosity when $t\sim t_{\rm shock}$.
As the accreted mass accumulates above the PNS magnetosphere, both the gas density and temperature gradually increase.
Consequently, the abrupt increase in $L^{\rm OBS}_{\bar{\nu}_{\rm e}}$ occurs much later than the shock formation time.
In other words, the time delay represents a waiting time until the gas density and temperature become high enough for neutrino emission to be significant.
To investigate its parameter dependence, we introduce the time at which the fallback luminosity reaches half of its maximum value, $t_{L/2}$.
We define the delay time scale as $t_{\rm delay}=t_{L/2}-t_{\rm shock}$.
As explained in Appendix \ref{sec:t-dependence}, $t_{\rm delay}$ is an increasing function of $B_{\rm 10km}$ and $r_{\rm PNS}$, while it is a decreasing function of $\dot{M}_{\rm fb}$.

\par
The middle panel of Figure \ref{fig:figure2} also includes an analytic solution of the neutrino light curve expected from the PNS cooling \citep[dashed lines,][]{Suwa2021}.
This solution provides the neutrino luminosity $L_{\rm PNS}$ and the mean energy $E_{\rm PNS}$ as functions of time.
These quantities are consistent with those obtained from numerical simulations at times $\gtrsim 1~{\rm s}$ after the core bounce.
We adopt $M_{\rm PNS}=1.4~M_\odot$ and $r_{\rm PNS}=10~{\rm km}$, and take the remaining parameters from model {\tt 147S} in \citet{Suwa2021}.
Since this solution assumes equal luminosities for neutrinos and antineutrinos of all three flavors, the analytic luminosity does not depend on the mass ordering (see Equation (\ref{eq:MSW})).
The onset time of fallback accretion after the core bounce depends on the progenitor structure and the explosion mechanism (see the references in Section \ref{sec:BC}).
Since the neutrino emission from the PNS evolves as the PNS cools, the relative contribution of the fallback neutrinos to the observed signal depends on when the fallback accretion begins.
To examine this dependence, we define $t_{\rm fb,start}$ as the time, measured from the core bounce, at which the fallback accretion starts.
We assume that this time coincides with the moment of the shock formation, $t=t_{\rm shock}$.
Three typical cases are considered: $t_{\rm fb,start}=1$, $10$, and $100~{\rm s}$.
The analytic luminosity decreases with increasing $t_{\rm fb,start}$ as the PNS cools down through the neutrino emission.

\par
The fallback luminosities are lower than the PNS luminosity for $t_{\rm fb,start}\lesssim 10~{\rm s}$.
This result indicates that we cannot distinguish the fallback signal from the PNS cooling component based on luminosity alone for $t_{\rm fb,start}\lesssim 10~{\rm s}$.

\par
We also examine the time evolution of the mean neutrino energy.
The local mean neutrino energy emitted from the gas at a temperature $T$ is given by $\left<\varepsilon\right>_{\rm loc}=(F_3/F_2)kT=3.15kT$, where $F_l=\int^\infty_0 x^l(1+e^x)^{-1}dx$.
We calculate the mean energy of the observed electron antineutrinos as
\begin{equation}
\left\langle \epsilon_{\bar{\nu}_e} \right\rangle^{\rm OBS}
=
\left\{
\begin{array}{l}
\frac{
L_{\bar{\nu}_e}^{\rm OBS}
}{
4\pi\int
\left[\left(
\cos^2\theta_{12}
\dot{q}^{\rm eff}_{\rm e}
+\sin^2\theta_{12}
\dot{q}^{\rm eff}_{\rm X}
\right)/
\left\langle \epsilon \right\rangle_{\rm loc}\right]
r^2\,dr
} \\
\hfill \text{(normal ordering)} \\
\frac{
L_{\bar{\nu}_e}^{\rm OBS}
}{
4\pi\int
\left[
\dot{q}^{\rm eff}_{\rm X}/
\left\langle \epsilon \right\rangle_{\rm loc}
\right]
r^2dr
} \\
\hfill \text{(inverted ordering)}
\end{array}
\right.
\end{equation}

\par
The bottom panel of Figure \ref{fig:figure2} shows the time evolution of $\left\langle \epsilon_{\bar{\nu}_e} \right\rangle^{\rm OBS}$.
Since both the neutrino luminosity and mean energy increase with gas temperature, the time evolution of $\left\langle \epsilon_{\bar{\nu}_e} \right\rangle^{\rm OBS}$ is very similar to that of $L^{\rm OBS}_{\bar{\nu}_{\rm e}}$.
This panel also includes the analytic solution $E_{\rm PNS}$ for $t_{\rm fb,start}=1$, $10$, and $100~{\rm s}$.
For both mass orderings, we find $\left\langle \epsilon_{\bar{\nu}_e} \right\rangle^{\rm OBS}\approx E_{\rm PNS}$ for $t_{\rm fb,start}=1~{\rm s}$, whereas $\left\langle \epsilon_{\bar{\nu}_e} \right\rangle^{\rm OBS}> E_{\rm PNS}$ for $t_{\rm fb,start}\ge10~{\rm s}$.
These results indicate that the fallback signal will be observed as high-energy neutrinos when $t_{\rm fb,start}\gtrsim 10~{\rm s}$.

\subsection{Neutrino spectra\label{sec:spectra}}

\par
Section~\ref{sec:time_evol} shows that the fallback signal will be observed as high-energy neutrinos when $t_{\rm fb,start}\gtrsim 10~{\rm s}$.
We therefore investigate the energy spectra of the fallback neutrinos and examine how they depend on $B_{\rm 10km}$ and $\dot{M}_{\rm fb}$.
Two representative terrestrial neutrino detectors are considered: SK and Jiangmen Underground Neutrino Observatory (JUNO), assuming a supernova distance of $D=10~{\rm kpc}$.
We also qualitatively assess the detectability of the fallback signal, while a quantitative assessment is presented in Section~\ref{sec:HED}.
For these analyses, we use the fallback spectra time-averaged over the interval defined below.

\par
The time window for averaging the spectra is defined based on $t_{L/2}$.
The finding that $\left\langle \epsilon_{\bar{\nu}_e} \right\rangle^{\rm OBS}>E_{\rm PNS}$ for $t_{\rm fb,start} \geq 10~{\rm s}$ suggests that $t_{L/2}$ may be identified from the time-energy distribution of detected neutrino events.
We test this possibility by generating mock event samples with Monte-Carlo simulations based on our fiducial model and the PNS cooling model (see Appendix~\ref{sec:t_interval}).
The mock sample for $t_{\rm fb,start}=100~{\rm s}$ shows a clear excess of high-energy events near $t_{L/2}$, suggesting that $t_{L/2}$ can be identified in this case.
However, the energies of the PNS cooling neutrinos are comparable to those of the fallback neutrinos for $t_{\rm fb,start}=10~{\rm s}$, indicating that $t_{L/2}$ is harder to identify.
A reliable method to estimate $t_{L/2}$ for $t_{\rm fb,start}=10~{\rm s}$ requires a detailed analysis.
We leave this task for future work.
The present study assumes an idealized situation in which $t_{L/2}$ can be estimated from the observed time-energy distribution.
We adopt the time window $t=[t_{L/2},t_{L/2}+t_{\rm delay}]$ for averaging the fallback spectra.
The qualitative trends explained in the following remain unchanged even when a different averaging interval is used, as discussed in Appendix~\ref{sec:t_interval}.

\par
The procedure for evaluating the neutrino energy spectra is as follows.
We assume a Fermi-Dirac distribution with a vanishing chemical potential \citep[e.g.,][]{Suwa2009},
\begin{eqnarray}
    \frac{dP}{d\varepsilon}=
    \frac{2}{3\zeta(3)(kT)^3}
    \frac{\varepsilon^2}{\exp[\varepsilon/kT]+1},
    \label{eq:dpde}
\end{eqnarray}
where $\varepsilon$ is the neutrino energy.
This distribution function is normalized such that $\int^\infty_0(dP/d\varepsilon)d\varepsilon=1$.
Using Equation (\ref{eq:dpde}), the number luminosities per unit energy for $\bar{\nu}_{\rm e}$ and $\bar{\nu}_{\rm X}$ are written as
\begin{eqnarray}
    \frac{d\dot{\mathcal{N}}_{\bar{\nu}_{\rm e}}^{\rm SRC}}{d\varepsilon}
    &=&4\pi\int
    \frac{\dot{q}^{\rm eff}_{\rm e}}{\left<\varepsilon\right>_{\rm loc}}
    \frac{dP}{d\varepsilon}
    r^2dr,\\
    \frac{d\dot{\mathcal{N}}_{\bar{\nu}_{\rm X}}^{\rm SRC}}{d\varepsilon}
    &=&4\pi\int
    \frac{\dot{q}^{\rm eff}_{\rm X}}{\left<\varepsilon\right>_{\rm loc}}
    \frac{dP}{d\varepsilon}
    r^2dr,
\end{eqnarray}
respectively.
By considering the flavor conversion in Equation (\ref{eq:MSW}), we obtain the neutrino flux per unit energy measured at Earth, $d\dot{\mathcal{N}}_{\bar{\nu}_{\rm e}}^{\rm OBS}/d\varepsilon$.

\par
SK is a water Cherenkov detector that uses ultrapure water \citep{Fukuda2003}.
The main detection channel of SK is the inverse beta interaction,
\begin{eqnarray}
 {\bar{\nu}}_{\rm e}+p\rightarrow e^++n.
 \label{eq:IBD}
\end{eqnarray}
We assume a detector volume of $M_{\rm det,SK}=32.5~{\rm kton}$ for this estimation.
The event rate in SK per unit energy is given by \citep{Suwa2025},
\begin{eqnarray}
    \frac{d\dot{\mathcal{N}}^{\rm SK}}{d\varepsilon}=
    \frac{2}{18}
    \frac{M_{\rm det,SK}}{m_{\rm u}}
    \frac{1}{4\pi D^2}
    \frac{d\dot{\mathcal{N}}_{\bar{\nu}_{\rm e}}^{\rm OBS}}{d\varepsilon}
    \sigma_{\bar{\nu}_{\rm e}}(\varepsilon),
    \label{eq:SK}
\end{eqnarray}
where $\sigma_{\bar{\nu}_{\rm e}}=\sigma_0(\varepsilon/{\rm MeV})^2$ is the inverse beta decay cross-section with $\sigma_0=9.4\times10^{-44}~{\rm cm}^2$.

\par
JUNO is a liquid scintillator detector \citep{JUNO2016}.
Its main detection channel is the inverse beta interaction (Equation \ref{eq:IBD}).
We assume a detector mass of $M_{\rm det,JUNO}=20~{\rm kton}$ as a fiducial value in this estimation.
JUNO is a scaled-up version of KamLAND \citep{Suzuki2014}, whose fiducial mass and number of target protons are $M_{\rm det,Kam}=1~{\rm kton}$ and $N_{\rm target}=5.98\times10^{31}$, respectively.
The event rate for JUNO is described as \citep{Eizuka2021},
\begin{eqnarray}
    \frac{d\dot{\mathcal{N}}^{\rm JUNO}}{d\varepsilon}=
    \frac{M_{\rm det,JUNO}}{M_{\rm det,Kam}}
    \frac{N_{\rm target}}{4\pi D^2}
    \frac{d\dot{\mathcal{N}}_{\bar{\nu}_{\rm e}}^{\rm OBS}}{d\varepsilon}
    \sigma_{\bar{\nu}_{\rm e}}(\varepsilon).
    \label{eq:JUNO}
\end{eqnarray}
We note that Equations (\ref{eq:SK}) and (\ref{eq:JUNO}) produce identical spectral shapes, differing only by an overall normalization factor.

\par
In addition, we calculate the spectra predicted from the PNS cooling model to assess the detectability of the fallback signal.
This PNS spectrum is based on the analytic solution in \citet{Suwa2021} and is calculated using the PNS temperature $T_{\rm PNS}$, which is related to the mean energy as $E_{\rm PNS}=(F_3/F_2)kT_{\rm PNS}$.
We assume a Fermi-Dirac distribution with zero chemical potential:
\begin{eqnarray}
    \frac{d\dot{\mathcal{N}}_{\rm PNS}}{d\varepsilon}
    =
    \frac{L_{\rm PNS}}{F_3(kT_{\rm PNS})^4}
    \frac{\varepsilon^2}{\exp[\varepsilon/kT_{\rm PNS}]+1},
\end{eqnarray}
which is normalized such that $\dot{\mathcal{N}}_{\rm PNS}=\int(d\dot{\mathcal{N}}_{\rm PNS}/d\varepsilon)d\varepsilon=L_{\rm PNS}/E_{\rm PNS}$.

\begin{figure}[tb]
\centering
\includegraphics[width=\autofigwidth]{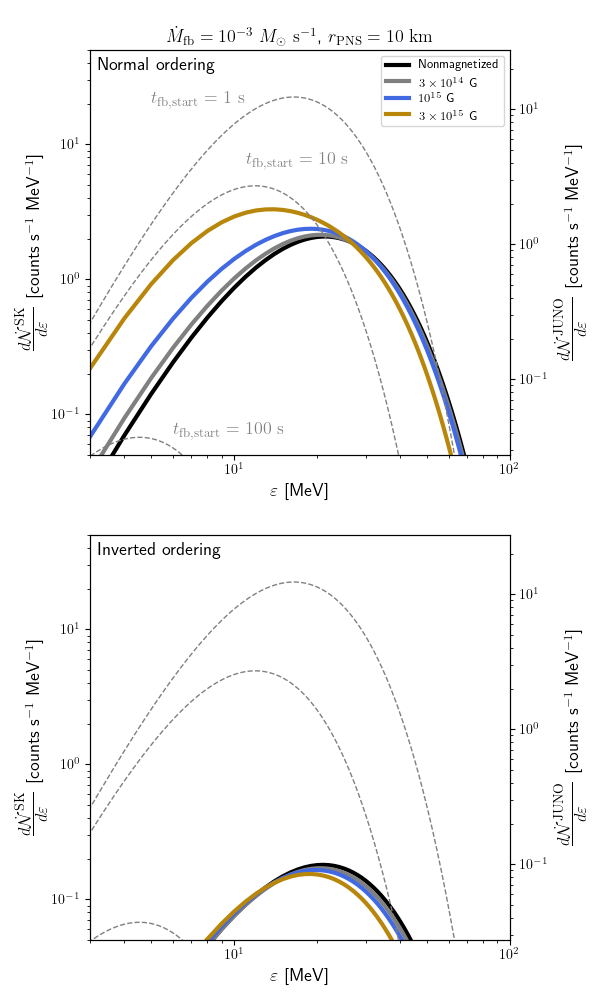}
\caption{
Neutrino spectra for different values of $B_{\rm 10km}$.
The left and right axes represent the event rates in SK and JUNO, respectively.
The spectra for the normal and inverted orderings are shown in the upper and lower panels, respectively.
Here, we adopt $\dot{M}_{\rm fb}=10^{-3}M_\odot~{\rm s}^{-1}$ and $r_{\rm PNS}=10~{\rm km}$.
Dashed lines represent the spectra estimated from the PNS cooling model.
The source distance is assumed to be $10~{\rm kpc}$.
Although the result for the non-magnetized case is very similar to that for $B_{\rm 10km}=3\times 10^{14}~{\rm G}$, a quantitative difference appears in the hardness-event-rate diagram, as discussed in Section \ref{sec:HED}.
\label{fig:figure3}}
\end{figure}

\par
Figure \ref{fig:figure3} presents the resulting spectra of the neutrino event rate for different values of $B_{\rm 10km}$ (solid lines).
We adopt $\dot{M}_{\rm fb}=10^{-3}~M_\odot~{\rm s}^{-1}$ and $r_{\rm PNS}=10~{\rm km}$.
The left and right axes represent the event rates for SK and JUNO, respectively.
The upper panel shows the spectra for the normal ordering, whereas the lower panel illustrates those for the inverted ordering.
The dashed lines represent the PNS spectra at $t=t_{\rm shock}$.
We consider three cases: $t_{\rm fb,start}=1$, $10$, and $100~{\rm s}$.
Since the PNS cools through neutrino emission, the PNS spectra become softer and the total event rate decreases as $t_{\rm fb,start}$ increases.

\par
For the normal ordering, the spectra become softer as $B_{\rm 10km}$ increases.
This spectral softening arises because a larger $r_{\rm M}$ leads to a lower $T$ there.
Although not shown in this panel, a lower $\dot{M}_{\rm fb}$ also produces a softer spectrum for the same reason.
The total event rate is $\mathcal{O}(10)-\mathcal{O}(10^2)$ for both detectors and is almost independent of $B_{\rm 10km}$, while it increases with increasing $\dot{M}_{\rm fb}$.
By comparing the resulting spectra with the PNS spectra, we find that the fallback signal appears as a high-energy tail at $\varepsilon\gtrsim 30~{\rm MeV}$ when $t_{\rm fb,start}=10~{\rm s}$.
Even in the strongly magnetized case ($B_{\rm 10km} \ge 3 \times 10^{15}~{\rm G}$), such high-energy neutrinos can be observed.
Therefore, the fallback signal is potentially detectable through the neutrino spectra when $\dot{M}_{\rm fb}\gtrsim 10^{-3}~M_\odot~{\rm s}^{-1}$ and $t_{\rm fb,start} \gtrsim 10~{\rm s}$.

\par
For the inverted ordering, the event rates are lower by an order of magnitude than those for the normal ordering.
As in the normal ordering, the spectra become softer with increasing $B_{\rm 10km}$ and decreasing $\dot{M}_{\rm fb}$.
The fallback signal is obscured by the PNS signal for $t_{\rm fb,start}\le 10~{\rm s}$.
Consequently, the signal will be observable in neither SK nor JUNO when $\dot{M}_{\rm fb}= 10^{-3}~M_\odot~{\rm s}^{-1}$ and $t_{\rm fb,start}\le 10~{\rm s}$.

\subsection{Hardness-event-rate diagram\label{sec:HED}}

\par
The spectral hardness of the fallback neutrinos depends on $B_{\rm 10km}$ and $\dot{M}_{\rm fb}$, suggesting that the fallback spectra can be used to diagnose these quantities.
We therefore introduce a quantitative measure of the hardness.
Specifically, we define a hardness ratio $H=\dot{\mathcal{N}}_{\rm high}/\dot{\mathcal{N}}_{\rm low}$, where $\dot{\mathcal{N}}_{\rm high}$ and $\dot{\mathcal{N}}_{\rm low}$ are the event rates of the fallback neutrinos integrated over energies above and below $30~{\rm MeV}$, respectively.
The energy threshold of $30~{\rm MeV}$ is chosen as a representative energy scale where the high-energy tail appears for $t_{\rm fb,start}=10~{\rm s}$.
We further introduce a hardness-event-rate diagram that relates $H$ to $\dot{\mathcal{N}}_{\rm high}$.
This diagram allows us to examine how the spectral properties depend on $B_{\rm 10km}$ and $\dot{M}_{\rm fb}$.
This diagram also enables a quantitative assessment of the detectability of fallback neutrinos and the applicability of our diagnostic.

\begin{figure}[tb]
\centering
\includegraphics[width=\autofigwidth]{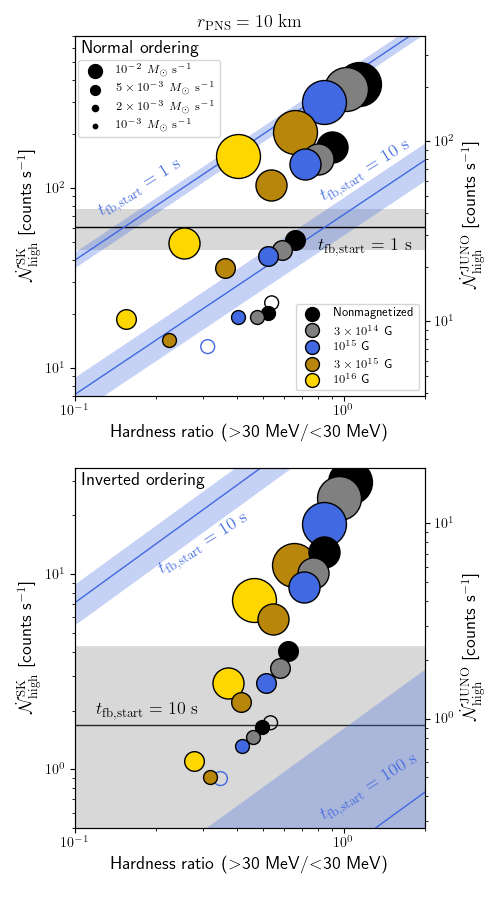}
\caption{
Hardness-event-rate diagram for $r_{\rm PNS}=10~{\rm km}$, assuming a source distance of $10~{\rm kpc}$.
The hardness ratio is defined as $H=\dot{\mathcal{N}}_{\rm high}/\dot{\mathcal{N}}_{\rm low}$, where $\dot{\mathcal{N}}_{\rm high}$ and $\dot{\mathcal{N}}_{\rm low}$ represent the event rate with $\varepsilon\ge 30~{\rm MeV}$ and $\varepsilon<30~{\rm MeV}$, respectively.
The left and right vertical axes denote $\dot{\mathcal{N}}_{\rm high}$ for SK and JUNO, respectively.
The black and blue lines represent $\dot{\mathcal{N}}_{\rm high}=\dot{\mathcal{N}}_{\rm high}^{\rm PNS}$ and $\dot{\mathcal{N}}_{\rm high}=\dot{\mathcal{N}}_{\rm low}^{\rm PNS}H$, respectively.
We adopt $t_{\rm fb,start}=1$, $10$, and $100~{\rm s}$ for these lines.
The gray and blue shaded regions indicate $2\sigma$ Poisson uncertainties, $\dot{\mathcal{N}}_{\rm high}^{\rm PNS}\pm2\sigma_{\rm high}$ and $(\dot{\mathcal{N}}_{\rm low}^{\rm PNS}\pm2\sigma_{\rm low})H$, respectively.
An observation time of $T_{\rm obs}=1~{\rm s}$ is assumed for estimating $\sigma_{\rm high}$ and $\sigma_{\rm low}$.
Open circles show the results in the two-dimensional models for $B_{\rm 10km}=0$ (black) and $B_{\rm 10km}=10^{15}~{\rm G}$ (blue) (see Section \ref{sec:multiD}).
\label{fig:figure4}}
\end{figure}

\par
Figure \ref{fig:figure4} presents the hardness-event-rate diagram.
The upper and lower panels show the results for the normal and inverted orderings, respectively.
The left and right vertical axes correspond to $\dot{\mathcal{N}}_{\rm high}$ for SK and JUNO, respectively.
The larger markers represent higher $\dot{M}_{\rm fb}$.
Since the spectra are softer for a stronger $B_{\rm 10km}$ and a lower $\dot{M}_{\rm fb}$ (Section \ref{sec:spectra}), both $\dot{\mathcal{N}}_{\rm high}$ and $\dot{\mathcal{N}}_{\rm high}/\dot{\mathcal{N}}_{\rm low}$ decrease with increasing $B_{\rm 10km}$ and decreasing $\dot{M}_{\rm fb}$.

\par
We quantitatively assess the detectability of the fallback signal by comparing $\dot{\mathcal{N}}_{\rm high}$ with that expected from PNS cooling, $\dot{\mathcal{N}}^{\rm PNS}_{\rm high}$.
Assuming an observation time of $T_{\rm obs}=1~{\rm s}$, we calculate the Poisson uncertainty as $\sigma_{\rm high}=\sqrt{\dot{\mathcal{N}}^{\rm PNS}_{\rm high}/T_{\rm obs}}$.
We consider that the signal is detectable when the detectability conditions, $\dot{\mathcal{N}}_{\rm high}>\dot{\mathcal{N}}^{\rm PNS}_{\rm high}+2\sigma_{\rm high}$ and $\dot{\mathcal{N}}_{\rm high}>1$, are satisfied.
The former condition means that the fallback signal with $\epsilon>30~{\rm MeV}$ exceeds the PNS signal with $\epsilon>30~{\rm MeV}$ by more than the $2\sigma$ Poisson fluctuation.
The black lines in Figure \ref{fig:figure4} show $\dot{\mathcal{N}}_{\rm high}=\dot{\mathcal{N}}^{\rm PNS}_{\rm high}$ for $t_{\rm fb,start}=1$ and $10~{\rm s}$.
The gray shaded regions indicate the $2\sigma$ Poisson uncertainty in $\dot{\mathcal{N}}^{\rm PNS}_{\rm high}$.

\par
For the normal ordering, the detectability conditions are satisfied in all models when $t_{\rm fb,start}=10~{\rm s}$.
The resulting $\dot{\mathcal{N}}_{\rm high}$ is greater than $10~{\rm counts~s^{-1}}$.
This result indicates that the fallback neutrinos are potentially detectable when $\dot{M}_{\rm fb}\gtrsim10^{-3}~M_\odot~{\rm s}^{-1}$ and $t_{\rm fb,start}\gtrsim10~{\rm s}$.
However, when $t_{\rm fb,start}=1~{\rm s}$, the detectability conditions are satisfied only for models with $\dot{M}_{\rm fb}=10^{-2}~M_\odot~{\rm s}^{-1}$ and with $\dot{M}_{\rm fb}=5\times10^{-3}~M_\odot~{\rm s}^{-1}$ if $B_{\rm 10km}\le3\times10^{15}~{\rm G}$.
Therefore, the detectability of the fallback signal is limited when the fallback phase begins at early times.

\par
For the inverted ordering, the parameter range in which the fallback signal is detectable is more limited than for the normal ordering.
For fixed $B_{\rm 10km}$ and $\dot{M}_{\rm fb}$, $\dot{\mathcal{N}}_{\rm high}$ is an order of magnitude lower than that for the normal ordering.
As a result, the detectability conditions are satisfied only by models with $\dot{M}_{\rm fb}\ge5\times10^{-3}~M_\odot~{\rm s}^{-1}$ when $t_{\rm fb,start}=10~{\rm s}$, except for model {\tt B1e16M5em3R10}.

\par
To assess the applicability of our diagnostic, we further compare $\dot{\mathcal{N}}_{\rm low}$ with that expected from the PNS cooling model, $\dot{\mathcal{N}}_{\rm low}^{\rm PNS}$.
The corresponding Poisson uncertainty is given by $\sigma_{\rm low}=\sqrt{\dot{\mathcal{N}}_{\rm low}^{\rm PNS}/T_{\rm obs}}$.
Since both $\dot{\mathcal{N}}_{\rm high}$ and $\dot{\mathcal{N}}_{\rm low}$ are needed to obtain the hardness ratio, we consider that our diagnostic is applicable when $\dot{\mathcal{N}}_{\rm low}>\dot{\mathcal{N}}_{\rm low}^{\rm PNS}+2\sigma_{\rm low}$ and $\dot{\mathcal{N}}_{\rm low}>1$ are satisfied in addition to the detectability conditions.
The blue lines in Figure \ref{fig:figure4} show $\dot{\mathcal{N}}_{\rm high}=\dot{\mathcal{N}}_{\rm low}^{\rm PNS}H$ for $t_{\rm fb,start}=1$, $10$, and $100~{\rm s}$.
The blue shaded regions indicate the $2\sigma$ Poisson uncertainty in $\dot{\mathcal{N}}_{\rm low}^{\rm PNS}$, $\dot{\mathcal{N}}_{\rm high}=(\dot{\mathcal{N}}_{\rm low}^{\rm PNS}\pm2\sigma_{\rm low})H$.
From the definition $H=\dot{\mathcal{N}}_{\rm high}/\dot{\mathcal{N}}_{\rm low}$, the region $\dot{\mathcal{N}}_{\rm high}>(\dot{\mathcal{N}}_{\rm low}^{\rm PNS}+2\sigma_{\rm low})H$ corresponds to the region $\dot{\mathcal{N}}_{\rm low}>\dot{\mathcal{N}}_{\rm low}^{\rm PNS}+2\sigma_{\rm low}$.

\par
In both mass orderings, within the range plotted in Figure \ref{fig:figure4}, the parameter range over which our diagnostic is applicable is narrower than that over which the fallback signal is detectable.
For the normal ordering, all the conditions are satisfied for models with $\dot{M}_{\rm fb}\ge5\times 10^{-3}~M_\odot~{\rm s}^{-1}$ and those with $\dot{M}_{\rm fb}=2\times 10^{-3}~M_\odot~{\rm s}^{-1}$ if $B_{\rm 10km}\ge3\times 10^{15}~{\rm G}$ when $t_{\rm fb,start}=10~{\rm s}$.
This result suggests that the hardness–event-rate diagram serves as a diagnostic of $B_{\rm 10km}$ and $\dot{M}_{\rm fb}$ in these parameter regions.
However, when $t_{\rm fb,start}=1~{\rm s}$, none of the models satisfy all the conditions.
For the inverted ordering, none of the models satisfy all the conditions when $t_{\rm fb,start}\le10~{\rm s}$.
The limitations on detectability and applicability explained here will be mitigated by future neutrino detectors, as discussed in Section \ref{sec:implications}.

\par
The statistical assessment presented above focuses on whether the fallback neutrino signal can be distinguished from the PNS-cooling component in two energy ranges, $\varepsilon\ge30~{\rm MeV}$ and $\varepsilon<30~{\rm MeV}$.
However, these criteria do not directly quantify the statistical uncertainty in the hardness ratio itself or the ability to distinguish models with different PNS magnetic field strengths.
A more detailed statistical analysis accounting for the Poisson fluctuations of both the fallback and PNS-cooling neutrino events is necessary to quantitatively constrain the PNS magnetic field.
We leave such an analysis for future work.

{\subsection{Dependence on the PNS radius}\label{sec:PNS_radius}}

\par
We have shown the results for $r_{\rm PNS}=10~{\rm km}$ in Sections \ref{sec:time_evol} - \ref{sec:HED}.
To investigate how the temporal and spectral properties depend on the compactness of the PNS, we study the cases with $r_{\rm PNS}=13$, $16$, and $20~{\rm km}$.
We adopt $\dot{M}_{\rm fb}=10^{-3}~M_\odot~{\rm s}^{-1}$.

\par
The models with larger $r_{\rm PNS}$ show distinct temporal behavior.
In some models with $r_{\rm PNS}>10~{\rm km}$, the shock radius undergoes quasi-periodic oscillations accompanied by variations in the neutrino luminosity (see Appendix \ref{sec:osc}).
We denote the corresponding oscillation timescale by $t_{\rm osc}$.
We find that $t_{\rm osc}$ increases with $r_{\rm PNS}$, suggesting that these oscillations may provide a probe of the PNS compactness.
To investigate the origin of these oscillations, we estimate the sound-crossing timescale between the PNS surface and the shock radius, $t_{\rm sound}$.
We find $t_{\rm osc}\gg t_{\rm sound}$, suggesting that these oscillations are not acoustic in origin.
Instead, we confirm that $t_{\rm osc}$ is comparable to the thermal timescale, suggesting that the oscillations are associated with the energy balance between accretion heating and neutrino cooling.
A detailed investigation of these oscillations will be presented in a subsequent paper.
We here focus on the dependence of the neutrino spectra on $r_{\rm PNS}$.

\begin{figure}[tb]
\centering
\includegraphics[width=\autofigwidth]{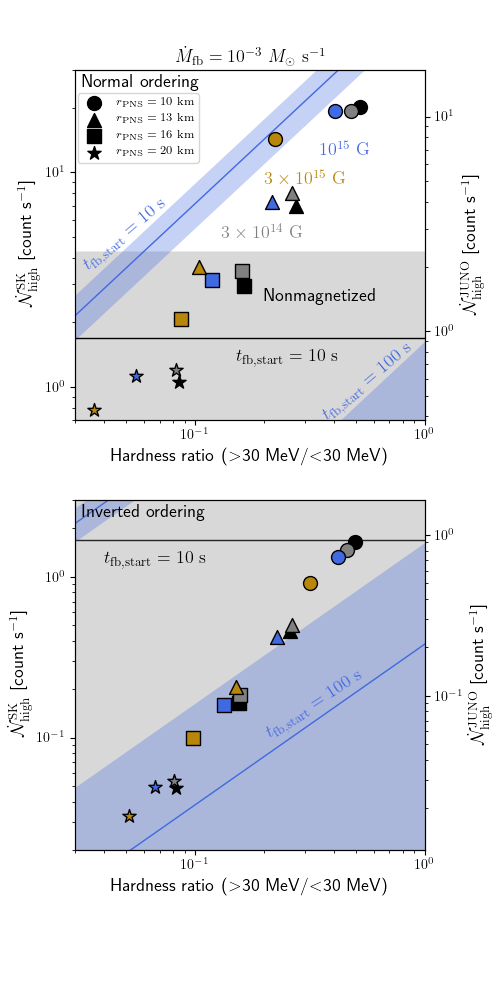}
\caption{
Same as Figure \ref{fig:figure4}, but for different values of $r_{\rm PNS}$.
Here, we adopt $\dot{M}_{\rm fb}=10^{-3}~M_\odot~{\rm s}^{-1}$.
\label{fig:figure5}}
\end{figure}

\par
Figure~\ref{fig:figure5} shows the hardness-event-rate diagram for different values of $r_{\rm PNS}$.
The upper and lower panels show the results for the normal and inverted orderings, respectively.
In both mass orderings, $\dot{\mathcal{N}}_{\rm high}$ and $\dot{\mathcal{N}}_{\rm high}/\dot{\mathcal{N}}_{\rm low}$ decrease as $r_{\rm PNS}$ increases.
In our simulations, the magnetospheric radius lies very close to the PNS surface (see Section \ref{sec:time_evol}).
Since $r_{\rm M}>r_{\rm PNS}$ is always satisfied, a larger $r_{\rm PNS}$ results in a larger $r_{\rm M}$.
The larger $r_{\rm M}$ results in a lower $T$ there.
Therefore, the fallback spectra become softer for a larger $r_{\rm PNS}$.

\par
The trend explained here is qualitatively similar to the dependence on $B_{\rm 10km}$, where a stronger magnetic field also leads to softer spectra through an increase in $r_{\rm M}$ (see Sections \ref{sec:spectra} and \ref{sec:HED}).
Therefore, the effects of $r_{\rm PNS}$ and $B_{\rm 10km}$ are partially degenerate in the hardness-event-rate diagram.
Although we adopt $M_{\rm PNS}=1.4~M_\odot$ in the present study, $r_{\rm PNS}$ and $M_{\rm PNS}$ are not independent in a realistic PNS.
The dependence on $r_{\rm PNS}$ suggests that uncertainties in the PNS compactness $M_{\rm PNS}/r_{\rm PNS}$ can affect our diagnostic.
Independent constraints on the PNS compactness are required to disentangle these effects, which will be discussed in Section \ref{sec:implications}.

{\section{Discussion}\label{sec:discussion}}

{\subsection{Observational implications}\label{sec:implications}}

\par
Our simulations suggest that the spectral hardness of fallback neutrinos can provide a new probe of the PNS magnetic field.
The hardness–event-rate diagram introduced in Section \ref{sec:HED} identifies the parameter range in which the fallback signal is detectable and the PNS magnetic field can be constrained.
The applicability of our diagnostic depends on the PNS magnetic field strength, the fallback accretion rate, the onset time of the fallback accretion, and the neutrino mass ordering.
In addition, we find that increasing either $r_{\rm PNS}$ or $B_{10\rm km}$ has a similar effect on the neutrino spectra.
This degeneracy prevents us from independently determining the PNS radius and magnetic field.
We therefore discuss how future observations may extend the applicability of our diagnostic and improve the constraints on the PNS magnetic field.

\par
Multimessenger observations will be important to break the degeneracy between the PNS compactness and magnetic fields in our diagnostic.
Although the present simulations have investigated the dependence on $r_{\rm PNS}$ at a fixed $M_{\rm PNS}=1.4~M_\odot$, the mass and radius of the PNS are not independent in a realistic PNS.
The dependence on $r_{\rm PNS}$ suggests that uncertainties in the PNS compactness can affect our diagnostic.
The PNS magnetic field cannot be independently determined from the proposed diagnostic unless the PNS compactness is constrained independently.
Gravitational wave observations provide a promising way to obtain such an independent constraint.
In core-collapse supernovae, post-bounce gravitational waves are expected to be generated mainly by oscillation modes of the newly formed PNS and its surrounding accretion flow.
The characteristic mode frequencies depend on the PNS compactness \citep[e.g.,][]{Sotani2016,Sotani2017,Torres2018,Torres2019a,Torres2019b,Abdikamalov2020,Bizouard2021}.
Therefore, simultaneous detection of fallback neutrinos and gravitational waves from a galactic supernova will help break the degeneracy and improve the constraint on the PNS magnetic field.

\par
The magnetic field inferred using the proposed diagnostic represents the field during the PNS phase and may differ from the NS magnetic field at later stages because of subsequent field evolution.
The fallback material can submerge the PNS field beneath the newly formed crust, potentially leading to the formation of CCOs \citep[e.g.,][]{Young1995,Muslimov1995,Geppert1999,Bernal2010,Ho2011,Shabaltas2012,Vigano2012,Torres2016,Shigeyama2018,Zhong2021}.
In our previous work \citep{Inoue2026}, we showed that field submergence can occur for $M_{\rm fb}\gtrsim 10^{-6}~M_\odot$ when $(B_{\rm 10km},r_{\rm PNS})=(10^{14}~{\rm G},10~{\rm km})$.
The detection of the fallback neutrinos by SK or JUNO requires $\dot{M}_{\rm fb} \gtrsim 10^{-3}~M_\odot\ {\rm s}^{-1}$, indicating that this mass condition is reached for $t_{\rm fb}\gtrsim 10^{-3}~{\rm s}$.
Since the typical fallback timescales are much longer, $t_{\rm fb}\sim 10^0-10^3~{\rm s}$, the detection of the fallback neutrinos may be consistent with the formation scenario of CCOs.
If the fallback neutrinos are detected from a substantial fraction of the galactic supernovae, this may suggest that CCOs are more common among young isolated NSs than inferred from the currently observed population \citep{deLuca2008,Torres2016}.

\par
The time evolution of the fallback neutrino spectra offers a potential probe of PNS magnetic-field evolution.
If the field submergence begins during the fallback phase, then the magnetic field strength at the PNS surface decreases over time.
The resulting reduction in the surface magnetic field reduces the magnetospheric radius and hardens the fallback spectrum.
Alternatively, the fallback accretion may amplify the PNS magnetic field through the Tayler-Spruit dynamo \citep{Barrere2022}, increasing the magnetospheric radius and softening the spectrum.
Time-resolved observations of the fallback neutrino spectra may therefore help distinguish between field submergence and amplification scenarios.

\par
Combining neutrino observations with electromagnetic observations will provide a direct way to test the subsequent evolution of the magnetic field.
After the surrounding material becomes optically thin, X-ray and radio observations can independently infer the dipole magnetic field of the young NS.
By comparing this late-time field with the PNS magnetic field inferred from the fallback neutrinos, we can investigate how the magnetic field evolves after birth and how the diversity of young NSs is established.
\footnote{Recent numerical simulations showed that the submerged field can be amplified by the dynamo process in the post-fallback phase \citep{Bambague2027}.
Combining the proposed diagnostic with late-time electromagnetic observations may help investigate such field amplification mechanisms.
}

\par
Hyper-Kamiokande (hereafter HK), an upgraded version of the SK, may extend the applicability of our diagnostic.
HK is under construction \citep{Hyper2018}.
Its reference volume is $M_{\rm det}=220~{\rm kton}$ and the event rate is simply scaled from that of SK.
The high statistics of HK may enable fitting of the full observed neutrino spectrum.
In Section \ref{sec:HED}, we have considered that our diagnostic is applicable when both the high- and low-energy event rates are distinguishable from the PNS-cooling component at the $2\sigma$ level and exceed $1~{\rm count~s^{-1}}$.
However, such a fitting may allow us to obtain the fallback spectrum even when $\dot{\mathcal{N}}_{\rm low}$ does not satisfy the $2\sigma$ criterion.
A quantitative assessment of this possibility requires a dedicated spectral-fitting analysis and is left for future work.

\par
The limitation associated with the neutrino mass ordering will be mitigated by future electron-neutrino observations.
The present study has focused on the electron-antineutrino signal, which has less diagnostic power for the inverted ordering than for the normal ordering.
However, the electron-neutrino signal is enhanced in the inverted ordering.
The Deep Underground Neutrino Experiment (DUNE) is a next-generation neutrino detector that uses liquid argon as its detection medium \citep{Abi2021}.
The detector's primary channel is the charged-current interaction of electron neutrinos with argon: $\nu_e + {}^{40}{\rm Ar} \rightarrow e^- + {}^{40}{\rm K}^{*}$.
Therefore, DUNE observations may make the diagnostic applicable in the inverted ordering for $\dot{M}_{\rm fb} \sim 10^{-3}~M_\odot\ {\rm s}^{-1}$ and $t_{\rm fb,start} \sim 10~{\rm s}$, for which the diagnostic with SK and JUNO is not applicable \citep{Akaho2024}.

{\subsection{Multidimensional effects}\label{sec:multiD}}

\par
Since we have adopted a one-dimensional approximation in this work, multidimensional effects are ignored.
The surface at $r=r_{\rm M}$ can be unstable to the magnetic Rayleigh–Taylor instability \citep[e.g.,][]{Kulkarni2008,Takasao2022,Das2024,Parfrey2024}.
This instability leads to the formation of turbulence, which may affect the neutrino spectra.
We therefore assess the impact of this instability.

\par
We briefly summarize the properties of the magnetic Rayleigh–Taylor instability.
This instability occurs in a magnetized stratified medium when a dense layer is supported by a lighter one under gravity \citep[e.g.,][]{Cumming2001,Stone2007}.
The behavior of the instability depends on the orientation of the perturbation wave vector $\vec{k}$ relative to the magnetic field vector $\vec{B}$.
For $\vec{k}\parallel\vec{B}$, magnetic tension acts as a restoring force, leading to stabilization at sufficiently short wavelengths.
The gravitational acceleration is written as $g=M_{\rm PNS}/r^2$.
From the balance between magnetic tension and gravity at the magnetospheric radius ($|\vec{B}|^2/4\pi\lambda_{\rm crit}=\rho g$), the critical wavelength at $r=r_{\rm M}$ is given by
$\lambda_{\rm crit}=[{|\vec{B}|^2}/({4\pi\rho})]({r_{\rm M}}/{r_{\rm g}})r_{\rm M}$.
On the other hand, the $\vec{k}\perp\vec{B}$ modes behave as in pure hydrodynamics and grow faster at shorter wavelengths.
The $\vec{k}\perp\vec{B}$ modes induce interchange motions that lead to deviations from the one-dimensional accretion structure at the magnetospheric boundary.
We examine the $\vec{k}\perp\vec{B}$ modes as a first step to assess the impact of turbulence on the neutrino spectra.

\par
For this purpose, we perform two-dimensional GRMHD simulations.
The simulation setup is summarized below.
We consider the fallback accretion on the $\theta=\pi/2$ plane.
The radius of the outer boundary is set to $r_{\rm out}=10^{3}~{\rm km}$.
In the azimuthal direction, the simulation domain is defined as $\phi=[4/9\pi,5/9\pi]$.
Periodic boundary conditions are applied in the $\phi$-direction.
The numerical grid consists of $(N_r, N_\theta, N_\phi) = (2048, 1, 256)$.
We initially impose a perturbation on the pressure by $0.1\%$ to break the symmetry.
We consider two cases with $B_{\rm 10km}=0$ and $10^{15}~{\rm G}$.
The remaining parameters in these models are the same as those in models {\tt BnM1em3R10} and {\tt B1e15M1em3R10}, respectively.
These simulations cannot take into account large-scale magnetic fields and magnetic tension.
We will discuss these effects in Section \ref{sec:limitations}.

\begin{figure*}[tb]
\centering
\includegraphics[width=\linewidth]{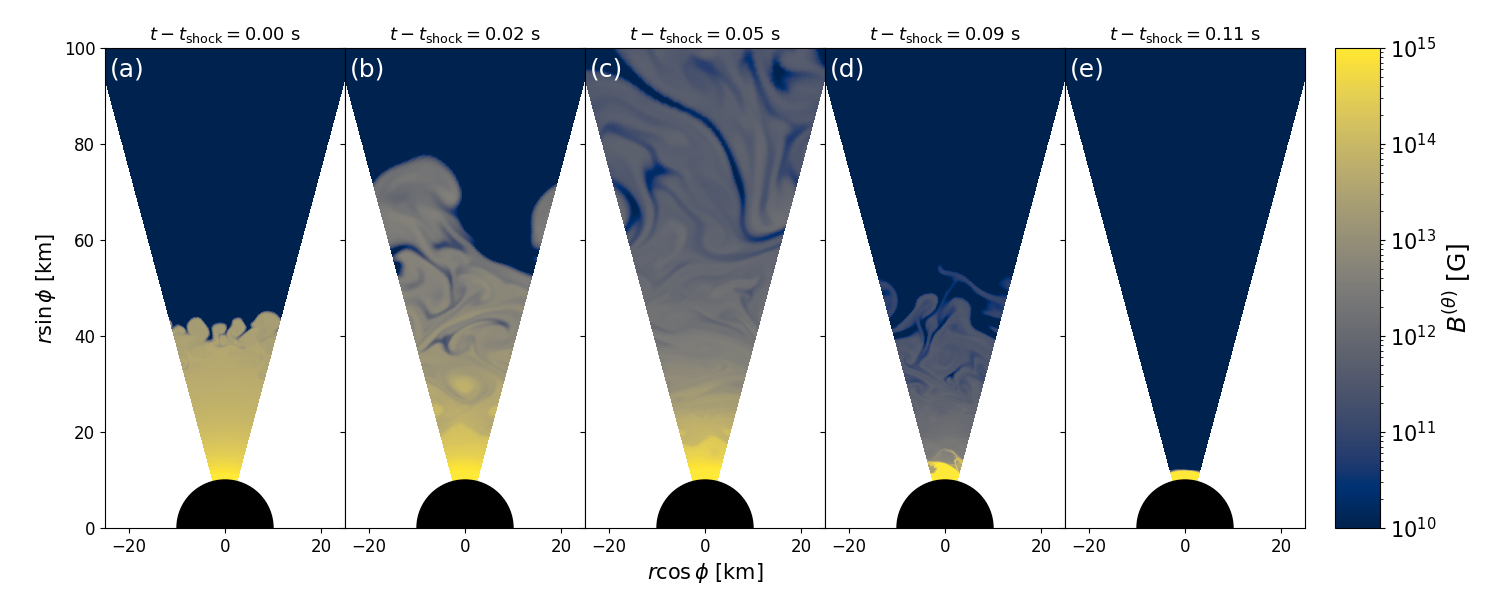}
\caption{
The resulting $B^{(\theta)}$ distributions in the two-dimensional simulation.
The system evolves from left to right.
The black circle represents the PNS surface.
\label{fig:figure6}}
\end{figure*}

\par
Figure \ref{fig:figure6} shows five snapshots of $B^{(\theta)}$ until the system reaches a quasi-steady state.
The system evolves from left to right.
The black circle represents the PNS.
Due to the growth of the $\vec{k}\perp\vec{B}$ modes, finger-like structures emerge when the shock forms (panel (a)).
The turbulence develops for $t-t_{\rm shock}\lesssim 0.05~{\rm s}$ (panels (b) and (c)).
However, it gradually decays once neutrino cooling becomes effective (panels (d) and (e)), consistent with the findings of \citet{Bernal2010,Bernal2013}.

\par
Using the azimuthally averaged cooling functions and temperature, we calculate the spectra in the same manner as in Section \ref{sec:spectra}.
The open circles in Figure \ref{fig:figure4} show the resulting relations between $\dot{\mathcal{N}}_{\rm high}/\dot{\mathcal{N}}_{\rm low}$ and $\dot{\mathcal{N}}_{\rm high}$ for the non-magnetized case (black) and for $B_{\rm 10km}=10^{15}~{\rm G}$ (blue).
Consistent with the trend explained in Section \ref{sec:HED}, $\dot{\mathcal{N}}_{\rm high}/\dot{\mathcal{N}}_{\rm low}$ and $\dot{\mathcal{N}}_{\rm high}$ for the non-magnetized case are greater than those for $B_{\rm 10km}=10^{15}~{\rm G}$.
This qualitative agreement justifies our use of the one-dimensional approximation for the broader parameter survey.
However, the quantitative differences between the one- and two-dimensional results indicate that multidimensional simulations are required to construct an accurate diagnostic.

In three dimensions, the $\vec{k}\parallel\vec{B}$ mode may develop and modify the nonlinear evolution of the turbulence.
To assess how the $\vec{k}\parallel\vec{B}$ mode affects the neutrino spectra, global three-dimensional simulations are needed; the $\vec{k}\parallel\vec{B}$ mode is relevant to the curvature of the global magnetic fields.
We leave the task for future work. 
The disappearance of the turbulence will occur even in the three-dimensional simulations because this process is driven by neutrino cooling \citep{Bernal2013}.

{\subsection{Model limitations}\label{sec:limitations}}

\par
In the limited two-dimensional models explored here, the qualitative trend with $B_{\rm 10km}$ remains unchanged.
However, there are multidimensional effects that are ignored in the present simulations.
We discuss these effects below.

\par
Realistic PNS magnetic-field geometries may modify the neutrino spectra.
The present simulations have neglected large-scale magnetic-field structures and magnetic tension.
However, these effects may lead to the formation of column-like accretion flows near the magnetic poles of the PNS \citep[e.g.,][]{Melatos2014,Takasao2022,Inoue2023}.
The resulting neutrino emission may consist of two spatially distinct components associated with the polar columns and the magnetospheric region.
The polar component may contain high-energy neutrinos with $\varepsilon>30~{\rm MeV}$ because the accretion energy can be dissipated within compact and hot regions \citep{Metzger2018}.
On the other hand, the magnetospheric component may be dominated by low-energy neutrinos.
A larger magnetospheric radius leads to a lower temperature there, which may reduce the contribution of the magnetospheric component to the overall emission.
If the contribution of the polar component remains significant, the resulting neutrino spectra may become harder with increasing PNS magnetic field strength.
This possibility should be further investigated by three-dimensional simulations.

\par
Although we have assumed a non-rotating PNS, the PNS rotation may also introduce important multidimensional effects.
When the PNS is rapidly rotating, the magnetospheric radius can exceed the corotation radius.
The accreted material is then expelled by the propeller effect \citep{Illarionov1975,Parfrey2017,Prasanna2026_1,Prasanna2026_2}.
Because the propeller mechanism primarily operates near the equatorial plane, neutrino emission near the magnetospheric radius may be suppressed.
In contrast, accretion through the polar columns may remain active.
As a result, the relative contribution of high-energy neutrinos may increase, leading to a harder neutrino spectrum.
The impact of the PNS rotation on the neutrino spectra should be investigated in future simulations.

\par
Magnetic reconnection may also modify the neutrino spectra during fallback accretion.
The reconnection may occur in small-scale current sheets generated by MHD turbulence, dissipating excess magnetic energy and heating the plasma locally.
The resulting local heating may enhance the high-energy component of the neutrino spectrum, although its impact remains uncertain.
A quantitative assessment of this effect requires resistive GRMHD simulations.

{\section{Summary and Conclusions}\label{sec:conclusion}}

\par
To understand the diversity of young isolated NSs, it is essential to observationally constrain the magnetic fields during the PNS phase. 
By performing one-dimensional GRMHD simulations of supernova fallback onto a magnetized PNS, we investigated whether fallback neutrinos can be used to diagnose the PNS magnetic fields.
Our simulations took neutrino cooling and the Helmholtz EoS into account, both of which are essential to accurately describe the dynamics of supernova fallback (Figure \ref{fig:figure1}).
We considered wide parameter ranges including the fallback mass accretion rate ($\dot{M}_{\rm fb}=10^{-3}$--$10^{-2}~M_\odot~{\rm s}^{-1}$), the magnetic field strength at $r=10~{\rm km}$ ($B_{\rm 10km}=10^{14}$--$10^{16}~{\rm G}$ and $B_{\rm 10km}=0$ for comparison), and the PNS radius ($r_{\rm PNS}=10$, $13$, $16$, and $20~{\rm km}$).
We also performed two-dimensional GRMHD simulations to assess the impact of MHD turbulence.
Our findings are summarized below.

\par
The relative contribution of the fallback neutrinos to the observed signal depends on the onset time of fallback accretion.
The luminosity and mean energy of the fallback neutrinos increase abruptly once the neutrino emission becomes significant and remain nearly constant after the system reaches a quasi-steady state (Figure \ref{fig:figure2}).
We compared these quantities with those predicted by an analytic PNS cooling model for different values of $t_{\rm fb,start}$, where $t_{\rm fb,start}$ is the onset time of the fallback accretion measured from the core bounce.
The fallback luminosity is lower than the PNS luminosity when $t_{\rm fb,start}\lesssim 10~{\rm s}$, whereas the mean energy is greater than the PNS neutrino energy when $t_{\rm fb,start}\gtrsim 10~{\rm s}$.

\par
The fallback neutrino spectra become softer with increasing $B_{\rm 10km}$ and decreasing $\dot{M}_{\rm fb}$ (Figure \ref{fig:figure3}).
The fallback neutrinos are emitted primarily near the magnetospheric radius.
Stronger magnetic fields and lower accretion rates lead to a larger magnetospheric radius.
The larger radius results in a lower temperature there, thereby producing the spectral softening.

\par
To quantify this spectral softening, we introduced the hardness ratio $H=\dot{\mathcal{N}}_{\rm high}/\dot{\mathcal{N}}_{\rm low}$, where $\dot{\mathcal{N}}_{\rm high}$ and $\dot{\mathcal{N}}_{\rm low}$ are the event rates of the fallback neutrinos integrated over energies above and below $30~{\rm MeV}$, respectively.
We also introduced the hardness-event-rate diagram that relates $H$ and $\dot{\mathcal{N}}_{\rm high}$.
This diagram provides a quantitative diagnostic of the PNS magnetic field and allows us to quantitatively assess the detectability of the fallback signal (Figure \ref{fig:figure4}).
For a source distance of $10~{\rm kpc}$ and the normal ordering, the fallback signal can be detected by SK and JUNO when $\dot{M}_{\rm fb}\ge10^{-3}~M_\odot~{\rm s}^{-1}$ for $t_{\rm fb,start}\ge10~{\rm s}$.
The parameter range in which the signal is detectable becomes narrower as $t_{\rm fb,start}$ decreases and is further limited for the inverted ordering.

\par
Within the range plotted in Figure \ref{fig:figure4}, the parameter range in which our diagnostic is applicable is narrower than that in which the fallback signal is detectable.
For the normal ordering and $t_{\rm fb,start}=10~{\rm s}$, our diagnostic is applicable for $\dot{M}_{\rm fb}\ge5\times10^{-3}~M_\odot~{\rm s}^{-1}$ and also for $\dot{M}_{\rm fb}=2\times10^{-3}~M_\odot~{\rm s}^{-1}$ if $B_{\rm 10km}\ge3\times10^{15}~{\rm G}$.
The parameter range becomes narrower for smaller $t_{\rm fb,start}$ and for the inverted ordering.

\par
The fallback neutrino spectra become softer with increasing $r_{\rm PNS}$, leading to a degeneracy with $B_{\rm 10km}$ (Figure \ref{fig:figure5}).
Since $r_{\rm PNS}$ and $M_{\rm PNS}$ are not independent, the radius dependence found here suggests that uncertainties in the PNS mass-radius relation can affect our diagnostic.
An independent constraint on the PNS compactness would therefore enable us to better constrain the magnetic field.
Simultaneous observations of fallback neutrinos and gravitational waves from a galactic supernova may help break this degeneracy and improve constraints on the PNS magnetic field.

\par
We discussed the effects of MHD turbulence on the neutrino spectra based on the two-dimensional simulations.
The magnetic Rayleigh-Taylor instability drives MHD turbulence, but the turbulence decays once neutrino cooling becomes effective (Figure \ref{fig:figure6}).
As a result, the turbulence does not change the qualitative dependence of the neutrino spectra on $B_{\rm 10km}$ within the scope of our two-dimensional simulations.
These results support the use of one-dimensional models to investigate the qualitative parameter dependence.
However, the spectra obtained from the two-dimensional models differ quantitatively from those obtained from the one-dimensional models.
This quantitative difference indicates that multidimensional simulations are required to construct an accurate diagnostic.
Although more realistic magnetic-field geometries and PNS rotation may further modify the spectra (see Section \ref{sec:limitations}), the present framework provides a basis for future multidimensional studies.

\appendix

\section{Parameter dependence of the time delay\label{sec:t-dependence}}

\par
As demonstrated in Section \ref{sec:time_evol}, there is a time delay between the shock formation and the increase in the neutrino luminosity.
The time delay represents a waiting time until the gas density and temperature at $r\sim r_{\rm M}$ become high enough for neutrino emission to be significant.
We investigate the parameter dependence of this time delay.

\begin{figure*}[tb]
\centering
\includegraphics[width=\linewidth]{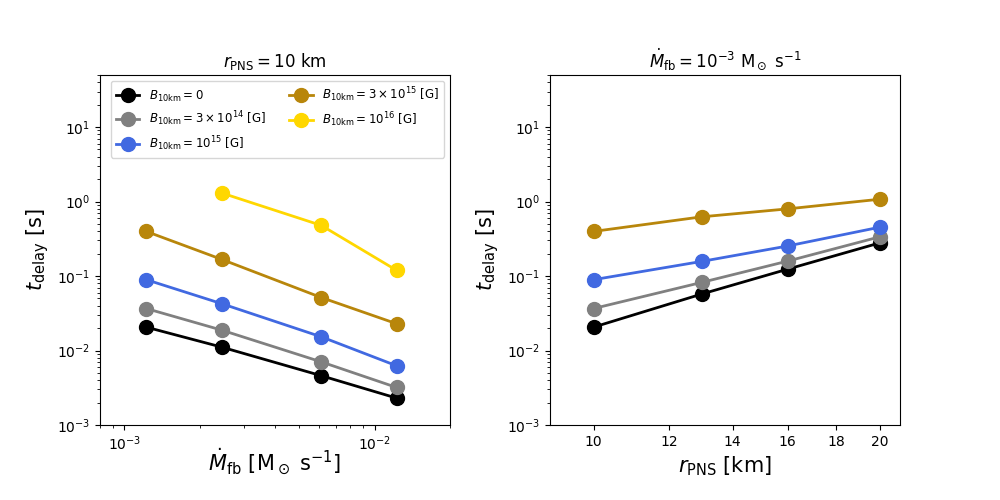}
\caption{
The dependence of the time delay on the fallback accretion rate (left) and the PNS radius (right) for different $B_{\rm 10km}$.
In the left and right panels, we adopt $r_{\rm PNS}=10~{\rm km}$ and $\dot{M}_{\rm fb}=10^{-3}~M_\odot~{\rm s}^{-1}$, respectively.
\label{fig:figure7}}
\end{figure*}

\par
Figure \ref{fig:figure7} shows $t_{\rm delay}$ as a function of $\dot{M}_{\rm fb}$ (left panel) and $r_{\rm PNS}$ (right panel) for different $B_{\rm 10km}$.
We adopt $r_{\rm PNS}=10~{\rm km}$ in the left panel, whereas $\dot{M}_{\rm fb}=10^{-3}~M_\odot~{\rm s}^{-1}$ is employed in the right panel.
The left panel shows that the delay decreases with increasing $\dot{M}_{\rm fb}$, while it increases with increasing $B_{\rm 10km}$.
These trends arise because a larger $r_{\rm M}$ leads to lower gas density and temperature there, resulting in a longer waiting time for neutrino emission.

\par
The right panel of Figure \ref{fig:figure7} shows that $t_{\rm delay}$ depends on $r_{\rm PNS}$.
We first focus on the non-magnetized case (black points).
The time delay increases with increasing $r_{\rm PNS}$.
This dependence arises because neutrino emission occurs mainly at $r\sim r_{\rm PNS}$.
The gravitational acceleration there decreases with increasing $r_{\rm PNS}$, resulting in lower density and temperature.
Consequently, a larger $r_{\rm PNS}$ leads to a longer $t_{\rm delay}$.

\par
For the weakly magnetized case ($B_{\rm 10km}\le10^{15}~{\rm G}$), the dependence on $r_{\rm PNS}$ is almost the same as that for the non-magnetized case because $r_{\rm M} \approx r_{\rm PNS}$ (see the top panel of Figure~\ref{fig:figure2}).
However, the dependence becomes weaker in the strongly magnetized case ($B_{\rm 10km}\ge3\times10^{15}~{\rm G}$).
This weak dependence arises because neutrino emission occurs mainly at $r\approx r_{\rm M}$.
Since the width of the magnetosphere increases with $B_{\rm 10km}$, the density and temperature at $r\approx r_{\rm M}$ are almost independent of $r_{\rm PNS}$ when $B_{\rm 10km}$ is strong.
Therefore, $t_{\rm delay}$ is almost independent of $r_{\rm PNS}$ for $B_{\rm 10km}\ge3\times10^{15}~{\rm G}$.

\par
The time evolution of the fallback signal may provide complementary information.
The fallback luminosity increases after $t_{\rm delay}$.
Therefore, the observed onset time of the fallback neutrinos, measured from the core bounce, is given by $t_{\rm fb,start}+t_{\rm delay}$.
Once $\dot{M}_{\rm fb}$ and $B_{10{\rm km}}$ are constrained from the spectrum, we can obtain $t_{\rm delay}$ and estimate the onset time of the fallback accretion.

\section{Mock observations and dependence of the spectra on the time-averaging interval\label{sec:t_interval}}

\par
In Section \ref{sec:time_evol}, we showed that $\left\langle \epsilon_{\bar{\nu}_e} \right\rangle^{\rm OBS}>E_{\rm PNS}$ for $t_{\rm fb,start} \geq 10~{\rm s}$.
This finding suggests that $t_{L/2}$ may be estimated from the time-energy distribution of the detected neutrino events.
We test this possibility by creating mock observation data for supernova neutrinos.
In addition, we examine the dependence of the spectra on the choice of time-averaging interval to assess the robustness of our results against uncertainties in estimating $t_{L/2}$.

To obtain the mock observation data, we use the Monte-Carlo simulation code {\tt FOREST} (Mori et al. in prep.).
\footnote{https://github.com/ForestRL/FOREST/releases/tag/v1.0.0}
This code generates individual detection events from an input time-dependent neutrino spectrum.
This code calculates the expected number and energy distribution of events in each time bin and randomly samples the detection time and energy of each event.
We assume a source distance of 10 kpc and the SK detector, and consider only the inverse beta decay channel given by Equation~(\ref{eq:IBD}).

\par
The input spectrum is constructed from our fiducial model {\tt B1e15M1em3R10} and the analytic solution of the PNS cooling (see Section \ref{sec:spectra}).
We use the numerical data in a time window of $t=[t_{\rm shock},t_{\rm L/2}+t_{\rm delay}]$.
Assuming that the onset time of the fallback accretion coincides with the moment of the shock formation, we obtain the total spectrum by summing their spectra.
We investigate two cases: $t_{\rm fb,start}=10~{\rm s}$ and $100~{\rm s}$.

\begin{figure}[tb]
\centering
\includegraphics[width=0.5\linewidth]{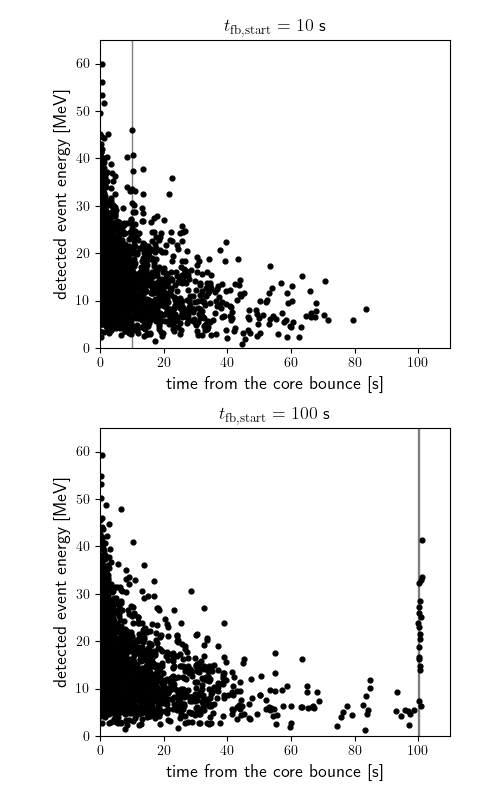}
\caption{
Mock data samples calculated from our fiducial model and PNS cooling model.
We assume a source distance of $10~\rm kpc$ and SK as the detector.
The narrow grey shaded regions represent the time window $t=[t_{\rm shock},t_{L/2}+t_{\rm delay}]$.
\label{fig:figure8}}
\end{figure}

\par
Figure \ref{fig:figure8} indicates the resulting mock data.
The upper and lower panels show the results for $t_{\rm fb,start}=10~{\rm s}$ and $100~{\rm s}$, respectively.
The region of $t=[t_{\rm shock},t_{L/2}+t_{\rm delay}]$ is represented by a narrow gray shaded region.
An excess of high-energy events appears around $t_{L/2}$ for $t_{\rm fb,start}=100~{\rm s}$.
On the other hand, the PNS cooling signal overlaps with the fallback signal for $t_{\rm fb,start}=10~{\rm s}$.
These results mean that $t_{L/2}$ can be estimated for $t_{\rm fb,start}=100~{\rm s}$, but it cannot be estimated for $t_{\rm fb,start}=10~{\rm s}$.

\begin{figure}[tb]
\centering
\includegraphics[width=0.5\linewidth]{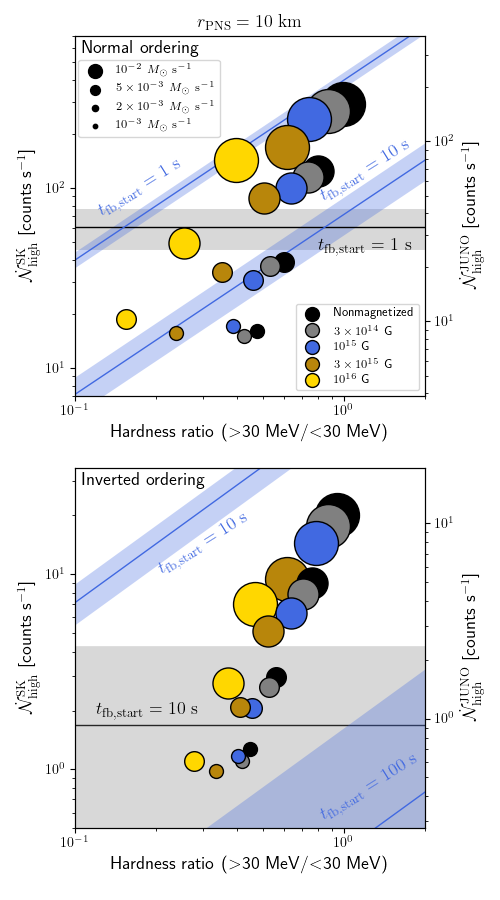}
\caption{
Same as Figure \ref{fig:figure4}, but based on spectra averaged over $t=[t_{L},t_{L}+t_{\rm delay}]$.
\label{fig:figure9}}
\end{figure}

\par
We discuss the dependence of the spectra on the choice of time-averaging interval.
We adopt a window of $t=[t_{L},t_{L}+t_{\rm delay}]$, where $t_{\rm L}$ means the time when the luminosity reaches its maximum value.
Figure \ref{fig:figure9} shows the hardness-event-rate diagram based on the spectra time-averaged over $t=[t_{L},t_{L}+t_{\rm delay}]$.
Although the results are slightly different from those in Figure \ref{fig:figure4}, the overall trends remain unchanged.
Therefore, we conclude that the main results are not affected by the choice of the time-averaging interval.

\section{Quasi-periodic variability for large PNS radii\label{sec:osc}}

\begin{figure}[tb]
\centering
\includegraphics[width=0.5\linewidth]{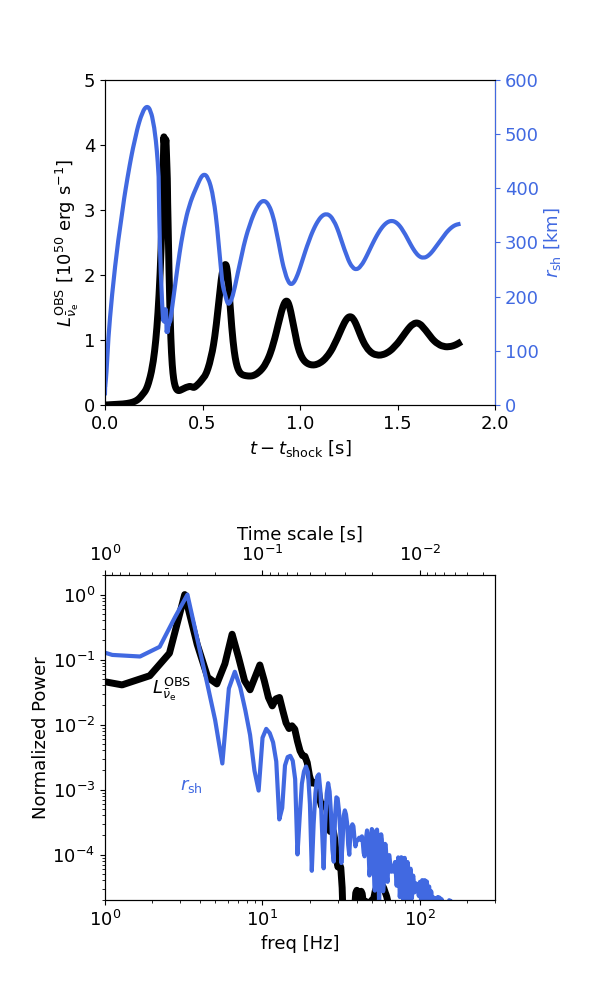}
\caption{
Time variability in model {\tt BnM1em3R20}.
The upper panel shows the time evolution of $L^{\rm OBS}_{\bar{\nu}_{\rm e}}$ (black) and $r_{\rm sh}$ (blue).
The lower panel shows their normalized power spectra.
Both quantities exhibit a characteristic variability with a timescale of
$t_{\rm osc}\sim0.3$--$0.4~{\rm s}$.
\label{fig:figure10}}
\end{figure}

\par
As mentioned in Section \ref{sec:PNS_radius}, some models with large $r_{\rm PNS}$ exhibit quasi-periodic oscillations in the shock radius and neutrino luminosity.
Figure \ref{fig:figure10} illustrates this behavior for model {\tt BnM1em3R20}.
The upper panel shows the time evolution of $r_{\rm sh}$ and $L^{\rm OBS}_{\bar{\nu}_{\rm e}}$.
After the shock formation, $r_{\rm sh}$ undergoes damped quasi-periodic oscillations.
The neutrino luminosity varies on a similar timescale, indicating that the luminosity variability is associated with the shock oscillation.

\par
To quantify the characteristic timescale of the variability, we calculate the power spectra of $r_{\rm sh}$ (blue) and $L^{\rm OBS}_{\bar{\nu}_{\rm e}}$ (black), as shown in the lower panel of Figure \ref{fig:figure10}.
Both power spectra have a prominent peak at a frequency of $f_{\rm osc}\sim2.5$--$3~{\rm Hz}$, corresponding to an oscillation timescale of $t_{\rm osc}\equiv f_{\rm osc}^{-1}\sim0.3$--$0.4~{\rm s}$.
We also find that $t_{\rm osc}$ increases with $r_{\rm PNS}$.
This dependence suggests that the temporal variability may provide complementary information on the PNS compactness.

\par
To understand the origin of these oscillations, we compare $t_{\rm osc}$ with the sound-crossing timescale in the post-shock region.
We estimate this timescale as
\begin{eqnarray}
    t_{\rm sound}
    =
    \int_{r_{\rm PNS}}^{r_{\rm sh}}
    \frac{dr}{c_{\rm s}}.
    \label{eq:tsound}
\end{eqnarray}
We find that $t_{\rm sound}$ is of order $10^{-3}~{\rm s}$ and is much shorter than $t_{\rm osc}$.
The relation $t_{\rm osc}\gg t_{\rm sound}$ suggests that the oscillation is not acoustic in origin.
Instead, we find that $t_{\rm osc}$ is comparable to the thermal timescale of the post-shock region.
This similarity suggests that the shock oscillation is associated with the energy balance between accretion heating and neutrino cooling.
A detailed investigation of the physical mechanism and observational implications of this variability will be presented in a subsequent paper.

\begin{acknowledgments}
This work was supported by JSPS KAKENHI grant Nos. JP24KJ0143, JP25K17439 (A.I.), JP24H02245, JP25K01035 (Y.S.), JP24K00632, JP26K17158 (R.A.), JP22KK0043, JP26K00769 (S.T.), 	JP25K00021, JP24K00668, JP23H04899 (K.K.), and JP24K00672, JP21H04488, JP24K00678 (H.R.T.).
A part of this research has been funded by the MEXT as "Program for Promoting Researches on the Supercomputer Fugaku" (Toward a unified view of the universe: from large-scale structures to  planets, JPMXP1020200109; A.I. and H.R.T.).
Numerical computations were performed with computational resources provided by HPE Cray XD2000  at the Center for Computational Astrophysics (CfCA) of the National Astronomical Observatory of Japan (NAOJ).
\end{acknowledgments}

\bibliographystyle{aasjournal}
\bibliography{library}

\end{CJK*}
\end{document}